\documentclass[onecolumn,noshowpacs,floatfix,11pt,nofootinbib]{revtex4}
\usepackage{amssymb,amsmath}
\usepackage[dvips]{graphics,color}
\usepackage[colorlinks,hyperindex]{hyperref}
\usepackage{epsfig}

\usepackage{float,stmaryrd,textcomp,bm,mathbbol}\usepackage{float,upgreek,yfonts}
\usepackage{graphicx,bigints}

\newcommand{\clt}{\textcolor{black}}
\newcommand{\ba}{\begin{eqnarray}}
\newcommand{\ea}{\end{eqnarray}}
\newcommand{\bege}{\begin{equation}}
\newcommand{\enge}{\end{equation}}

\newcommand{\beq}{\begin{eqnarray}}
\newcommand{\benu}{\begin{enumerate}}
\newcommand{\enu}{\end{enumerate}}
\newcommand{\eeq}{\end{eqnarray}}

\newcommand{\gbb}{AdS${}_4$ generalized extremal brane}
\newcommand{\gbbs}{AdS${}_4$ generalized extremal branes}
\newcommand{\adr}{AdS${}_4$-RN black brane}

\definecolor{green1}{RGB}{0,128,0} 
\hypersetup{backref=true,pagebackref=true}
\hypersetup{%
  colorlinks = true,
  linkcolor  = blue,
  citecolor = cyan,
}
\usepackage{bookmark,textgreek}
\usepackage{hyperref,color,xcolor}
\hypersetup{hidelinks,hyperindex=true,colorlinks=true,breaklinks=true,urlcolor= blue}
\hypersetup{%
  colorlinks = true,
  linkcolor  = blue
}\usepackage{amssymb}
\newsavebox{\foobox}

\usepackage{graphicx,float,tikz}
\usepackage[all]{xy}
\newcommand\ringring[1]{%
  {
   \mathop{\kern0pt #1}\limits^{
     \vbox to-1.85ex{
       \kern-2ex 
       \hbox to 0pt{\hss\normalfont\kern.1em \r{}\kern-.45em \r{}\hss}%
       \vss 
     }
   }
  }
}

\newcommand\orcidroldao{{\href{https://orcid.org/0000-0003-3978-532X}{\orcidicon}}}
\newcommand{\orcidicon}{%
	\begin{tikzpicture}
	\draw[lime, fill=lime] (0,0)
		circle [radius=0.16]
		node[white] {{\fontfamily{qag}\selectfont \tiny ID}};
	\draw[white, fill=white] (-0.0625,0.095)
		circle [radius=0.007];
	\end{tikzpicture}	\hspace{-2mm}
}

 \newcommand{\bea}{\begin{eqnarray}}
 \newcommand{\eea}{\end{eqnarray}}

\usepackage{hyperref}
\usepackage{xcolor}

\begin{document}

\title{Generalized \clt{extremal} branes in AdS/CMT and holographic superconductors}

\author{Roldao da Rocha\orcidroldao\!\!}
\email{roldao.rocha@ufabc.edu.br}
\affiliation{Federal University of ABC, Center of Mathematics, Santo Andr\'e, 09580-210, Brazil}


\begin{abstract}
\clt{AdS$_4$ generalized extremal branes} are scrutinized in the context of AdS/CFT.  Holographic superconductors are studied as dual objects to \clt{\gbbs}, whose coefficients of response and transport in the dual condensed matter theory are calculated and discussed. The holographic Weyl anomaly is also addressed. 
The holographic superconductor bound current is shown to be enhanced by the parameter controlling the family of AdS$_4$ generalized extremal branes, when compared to the standard AdS${}_4$-Reissner--Nordstr\"om black brane results. In the probe limit, the electrical DC conductivity for holographic superconductors with \clt{\gbb}\,\! dual background is also reported. 
 \end{abstract}
\maketitle

\section{Introduction}
The AdS/CFT correspondence consists of a solid apparatus where  
 strongly-coupled field theories are studied. Any given field theory, including 
 finite temperature ones, has a hydrodynamical description in the 
 infrared (IR) limit, corresponding to long-length scales. 
 In the anti-de Sitter (AdS) bulk space, a theory of gravity is dual to the conformal field theory (CFT) on the boundary. At finite temperature, the bulk geometry consists of an AdS black brane with an event horizon. The holographic duality conjectures that the CFT at the long-scale regime, on the boundary, must be ruled by the near-horizon limit in the bulk. For instance, any general relativistic  
black hole presents a spurious  fluid on its horizon, consisting of the so-called membrane paradigm, whose low-energy regime is a strongly-coupled
field theory \cite{malda,Iqbal:2008by,daRocha:2017cxu}.  
Transport coefficients can be computed by implementing  gravitational perturbations in the black brane horizon. The shear viscosity of the dual field theory can be read off the absorption cross-section of the graviton by the black brane  \cite{kss}. Einstein's equations in the AdS bulk correspond to the Navier--Stokes equations on the AdS boundary, constituting the fluid/gravity correspondence \cite{Bhattacharyya:2008jc}. 

AdS/CFT states that AdS--Schwarzschild black branes have the same metric as  
the stack of $N$ $D_3$ branes, which is dual to finite-temperature  
$\mathcal{N} = 4$ supersymmetric SU($N$) Yang--Mills theory, in the limit of large
$N$ and large ’t Hooft coupling \cite{malda}. A $D_3$ brane embedded in the AdS$_5$ bulk yields a bulk Riemann tensor related to the $D_3$ brane Riemann tensor by the Gauss--Codazzi equations. 
A constraint to Einstein's effective field equations  within the holographic membrane paradigm is to demand the general-relativistic limit, consisting of a rigid brane with infinite tension as a low-energy limit \cite{Ferreira-Martins:2019svk,daRocha:2020jdj,mgd2,DaRocha:2019fjr}. However, 
$D_3$ branes do have finite tension. 
AdS/CFT and the membrane paradigm were successfully studied in Ref. \cite{Antoniadis:1998ig}. The AdS/CFT fermionic sector has been tested in supersymmetric backgrounds \cite{Bonora:2014dfa}. However, a
static AdS black hole in the supersymmetric limit can have a naked singularity, which can be avoided by turning on rotation, as in the Reissner--Nordstr\"om spacetime. A precise link between braneworld scenarios and AdS/CFT duality has been explored, allowing black branes and their underlying hydrodynamics to be probed by well-known AdS/CFT methods  \cite{ads_memb}. 
The AdS$_4$-Reissner--Nordstr\"om (AdS$_4$-RN) black brane  plays a prominent role in this procedure, in the context of AdS/CMT  (condensed matter theory) correspondence,  for being dual to a finite temperature (2+1)-dimensional CFT, describing a conserved U(1) charge in the boundary \cite{Davison:2011uk,Cadoni:2009xm}. The boundary conditions at infinity, in the AdS$_4$ bulk, correspond to ultraviolet values of couplings in the field theory on the boundary. Several seminal results about AdS$_4$-RN black branes and the dual CMT  have been seamlessly implemented, in particular concerning strange metals and holographic superconductors  as well \cite{Iqbal:2011in,Giordano:2018bsf,Mohammed:2012gi}.

Superfluids and holographic superconductors can be described by a dual AdS$_4$-RN geometry coupled to an AdS$_4$ scalar field \cite{Hartnoll:2008vx,Gubser:2008px}. 
The local U(1)\;symmetry  of the AdS$_4$ bulk is associated
with a global U(1)\;symmetry on the AdS$_4$ boundary. A Higgs-like scalar field in the bulk corresponds to the condensate in the dual field theory \cite{Brihaye:2011vk}. 
In the AdS$_4$-RN black hole near-horizon region, the Abelian symmetry in AdS$_4$  can be spontaneously broken, since the scalar field effective mass can attain negative values, below the so-called Breitenlohner--Freedman bound \cite{Breitenlohner:1982jf}. It causes the scalar field to be unstable, undergoing  condensation into an atmosphere near the AdS$_4$-RN black hole horizon. 
The AdS$_4$--RN black brane is already known to describe strange metals in the holographic duality setup. Holography can relate it to finite-density, strongly-coupled, dual objects in the AdS$_4$ boundary, representing metallic states of matter at zero temperature \cite{Davison:2013txa}. These metallic states in condensed matter are observed experimentally in strongly-correlated systems, encompassing high-$T_c$ superconductors, and constitute a class of non-Fermi-liquids corresponding to holographic strange metals. The description of the simplest kinds of strange metals, through a weakly-coupled dual theory of  gravity, can be emulated by the charged AdS$_4$--RN black brane \cite{Hartnoll:2009ns}. Since the discovery of  superconductivity at high temperatures in cuprates, holographic superconductors in AdS/CMT have been playing a prominent role in describing them. Superconductivity phenomena in cuprates cannot be reported by the standard theory based on the Bardeen--Cooper--Schrieffer microscopic theory of superconductivity, being usually described by a different mechanism.

By promoting the standard AdS$_4$--RN black brane metric to AdS$_4$ generalized black branes, we expect to be able to describe a higher range of holographic superconductor materials. Hence  \clt{\gbbs}\,\! may model a larger class of strange metals, whose transport and response coefficients, such as the electrical conductivity, the shear viscosity-to-entropy density ratio, and the bulk viscosity-to-entropy density ratio can be fine-tuned by the parameter that governs the family of \clt{\gbb}. 
This paper is organized as follows: in Sec. \ref{sec:2} AdS$_4$ generalized black branes are introduced in the ADM formalism, whose metric can be obtained from the Hamiltonian and momentum constraints. The very existence of a second event Killing horizon yields the parameter $\beta$ that regulates the family of \clt{\gbbs}\,\! to depend on the U(1) charge. The Maxwell equations are solved, and the magnetic potential is obtained, having an appropriate formula involving the $\beta$-dependent chemical potential. Using the GKPW relation,  the shear viscosity-to-entropy density and the bulk viscosity-to-entropy density ratios are derived and discussed, yielding a more strict range for $\beta$. The conformal anomaly is also addressed. Sec. \ref{hs1ads} is devoted to approaching the holographic superconductor in the AdS$_4$ generalized black brane setup, described by a  complex charged scalar field emulating the Higgs field in Ginzburg--Landau theory, minimally coupled to Einstein--Maxwell theory. The Breitenlohner--Freedman bound
is then discussed, and the equations of motion are solved and analyzed. Type-I and type-II superconductors are addressed. For type-II superconductors, a vortex lattice, engendering a supercurrent, is investigated. The bound current will be shown to be enhanced when compared to the AdS$_4$-RN black brane case, whereas the supercurrent is not modified. In Sec. \ref{sec:4} the electrical DC conductivity of \clt{\gbbs}\,\! at vanishing temperature is also calculated for the dual holographic superconductors, and analyzed as a function of the frequency-to-chemical potential ratio. 
Finally,  concluding remarks and some perspectives for the forthcoming developments are presented in Sec. \ref{sec:5}.


\section{AdS$_4$ generalized black branes} \label{sec:2}

The AdS$_4$--Reissner--Nordstr\"om (AdS$_4$--RN) black brane background is a well-known solution of the Einstein--Maxwell-AdS$_4$ theory,  
\beq
S = \frac{1}{16\pi G_4} \int \left(R-2\Lambda_4 - \frac14 F_{\mu\nu}F^{\mu\nu}\right)\,\sqrt{-g} \,d^4x,\label{em1}
\eeq
with cosmological constant $\Lambda_4=-3/L^2$, where $R$ is the Ricci  scalar curvature,  and $F_{\mu\nu}$ is the Maxwell electromagnetic field strength  \cite{Davison:2011uk,Cadoni:2009xm}. The AdS$_4$--RN solution with a planar horizon has metric 
\begin{eqnarray}\label{rnads4}
ds_4^2=\frac{r^2}{L^2}\left(-f(r)dt^2+dx^2+dy^2\right) +\frac{L^2}{r^2\,f(r)}dr^2,
\end{eqnarray}
with \begin{eqnarray}
f(r)&=&1-(1+\mathfrak{q}^2)\left(\frac{r_0}{r}\right)^3+\mathfrak{q}^2\left(\frac{r_0}{r}\right)^4.\label{rnads41}
\end{eqnarray}
where $\mathfrak{q}$ denotes the black hole charge and $r_0$ represents the event horizon. \clt{The standard AdS$_4$--RN solution in Eqs. (\ref{rnads4}, \ref{rnads41}) is not valid
only in the extremal case, but also in the non-extremal case. The solution also has the electromagnetic field strength component $F_{tz} = \mathfrak{q}r_0/L^2$,
in Poincar\'e coordinates $u = r_0/r$. The extremal limit happens as  $\mathfrak{q}\to\pm\sqrt{3}$, with $r_0/L^2$ fixed.} In the gauge/gravity dictionary, the black brane planar horizon is consistent with a dual QFT on a (2+1)-Minkowski background and is useful for describing condensed matter systems. Also, the AdS$_4$--RN black brane charge corresponds to a dual field theory with a charge density encoding, e.g., the fluid flow of electrons in metals. 
The AdS$_4$--RN black brane is the unique regular solution of (\ref{em1}) with Poncar\'e symmetry, with electric flux at infinity. Expanding the solution (\ref{rnads4}, \ref{rnads41}) in the near-horizon yields the background topology AdS$_2 \times\mathbb{R}^2$ \cite{Hartnoll:2014gaa}.  The near-horizon extremal geometry is dual to a field theory in the infrared limit.

The AdS/CFT membrane paradigm can set the AdS$_4$ spacetime as the boundary of an AdS$_5$ bulk, with cosmological constant 
$\Lambda$, related to the boundary vacuum energy density.
The AdS$_5$ bulk, with electromagnetic field strength, satisfies Einstein's equations, 
\begin{eqnarray}
{R}_{AB}-\frac12 R\,g_{AB}=\Lambda\,g_{AB}+{\color{black}{T_{AB}}},
\label{Dm1eq}
\end{eqnarray}
for 
\begin{equation}
{\color{black}{T_{AB}=F_{AC}F_{B}^{\text{ }C}-\frac{1}{4}g_{AB}F^{2},}}
\end{equation}
{\color{black}{where $F^{2}=F_{AB}F^{AB}$.}}
Projecting Eq. \eqref{Dm1eq} onto AdS$_4$, and introducing Gaussian coordinates $(x^\mu,w$) in the AdS$_5$ bulk yields, at $w=0$:
\begin{eqnarray}
{R}_{\mu w}=0\ ,\qquad\qquad 
R=\Lambda_4.
\label{Deq}
\end{eqnarray}
For static solutions, Eqs.~(\ref{Deq}) encode the Hamiltonian and the ADM momentum constraints \cite{Casadio:2001jg} encompassing the embedding of the \adr\, on the AdS$_5$ bulk. To apply the ADM formalism to generate a family of \clt{\gbbs}, let us emulate the metric in Eq. \eqref{rnads4}, still considering the temporal component in Eq. \eqref{rnads41}, however with a radial component deforming $f(r)$, as  \cite{Ferreira-Martins:2019wym} 
\begin{eqnarray}\label{rngen}
ds_4^2=\frac{r^2}{L^2}\left(-f(r)dt^2+dx^2+dy^2\right) +\frac{L^2}{r^2} n(r) dr^2,
\end{eqnarray}
The $r$-coordinate can be identified to the renormalization group flow scale. A family of analytic solutions
of the form (\ref{rngen}) can be obtained by
relaxing the condition $f(r)=-1/n(r)$ in Eq. (\ref{rnads4}), by fixing $f(r)$ as
in Eq. (\ref{rnads41}) for the AdS$_4$-RN spacetime,  and finding the most general solutions for
the constraints (\ref{Deq}).
The momentum constraints are identically satisfied by
the metric (\ref{rngen}) and the Hamiltonian constraint
reads  
\beq
&&{n}(r) \left[\frac{(r\!-\!r_0){n}'(r)}{L^4} \left(-\mathfrak{q}^2 r_0^3\!+\!r^3\!+\!r^2 r_0\!+\!r r_0^2\right)
   \left(\left(\mathfrak{q}^2\!+\!1\right) r r_0^3\!-\!2 \mathfrak{q}^2 r_0^4\!+\!2 r^4\right)(r\!-\!(2 \beta  \!+\!7) r_0)^3\!+\!4 r^3\right]\nonumber\\&&+\frac{4 r^4{n}'(r)}{{n}(r)}\!-\!\frac{r^3}{\left[\left(\mathfrak{q}^2\!+\!1\right) r r_0^3\!-\!\mathfrak{q}^2 r_0^4\!-\!r^4\right]^2} \left[4 \mathfrak{q}^4 r_0^8\!-\!24 \left(\mathfrak{q}^2\!+\!1\right) r^5
   r_0^3\!+\!32 \mathfrak{q}^2 r^4 r_0^4\left[6 \beta \!+\!(4 \beta \!+\!59)
   \mathfrak{q}^2\!+\!21\right]\right.\nonumber\\&&\left.+3 \left(\mathfrak{q}^2+1\right)^2 r^2 r_0^6+\left(2 \beta -27 \mathfrak{q}^2-29\right)r^2r_0^2-8 \mathfrak{q}^2 \left(\mathfrak{q}^2+1\right) r r_0^7+12
   r^8\right]=0,\label{adm}\eeq
   where $\beta$ is a real arbitrary parameter that governs the deformation of the \adr. Hence, the solution 
\begin{equation}
	 n(r)=\frac{1}{f\left(r\right)}\left\{ \frac{1-\frac{r_{0}}{r}}{1-\frac{r_{0}}{r}\left[1+\frac{1}{3}\left(\beta-1\right)\right]}\right\} \ , \label{eq:1}
\end{equation}
satisfies Eq. (\ref{adm}), as long as the temporal component $f(r)$ is given by Eq. (\ref{rnads41}). 
Notice that for $\beta \to 1$, the metric Eq. \eqref{rngen} is precisely the AdS$_4$-RN black brane background. Similarly to the \adr\,, the \clt{\gbb}\,\! also describes the physics underlying the electric flux in an asymptotically AdS${}_4$ geometry. From now on, the value $L=1$ will be assumed.
\par
\begin{figure} [H]
	\centering
	\includegraphics[scale=0.55]{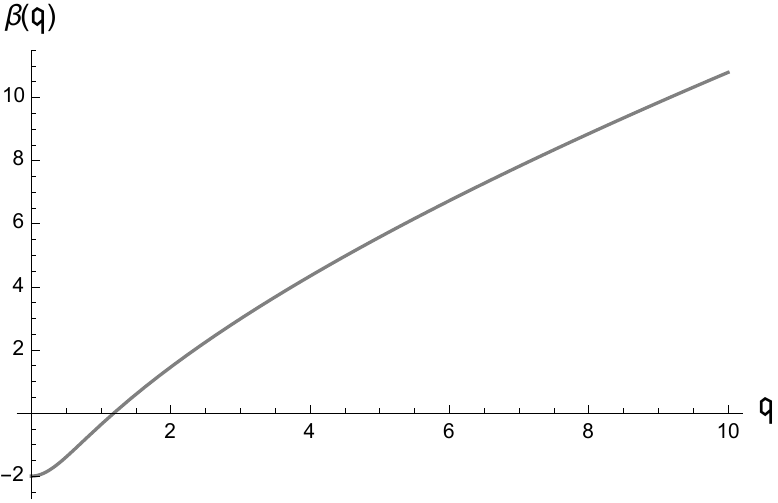}
	\caption{Plot of the parameter $\beta(\mathfrak{q})$ as a function of $\mathfrak{q}$, in Eq.  \eqref{eq:7}. }
		\label{dois}
\end{figure}

 The limit $\displaystyle\lim_{r\to r_{\beta}}1/n(r)=0$, 
in Eq.  \eqref{eq:1}, for the (coordinate) singularity 
\begin{equation} \label{eq:4}
{\color{black}{r_{\beta}=\frac{r_{0}}{3}\left[2+\beta\right]}}\,,
\end{equation}
is essential for studying event horizons in \clt{\gbbs}.  To realize whether the singularity $r_{\beta}$ is an event horizon, the Killing horizon condition $\upxi^{\mu}\upxi_{\mu}=0$, for $\upxi^{\mu}$ being the associated timelike Killing vector, yields two possible values for  $\beta$. The first one,  
	\begin{align}
	\beta&=1\ , \label{eq:6}	
		\end{align}
 corresponds to the 	standard AdS${}_4$--RN black brane. The other possibility that is consistent with the identification of $r_\beta$ with an event horizon is given by  \cite{Ferreira-Martins:2019wym}
	\begin{align}
	\color{black}\beta&\color{black}=\frac{2\sqrt[3]{2}}{\sqrt[3]{-7-27\mathfrak{q}^{2}+3\sqrt{3}\sqrt{3+14\mathfrak{q}^{2}+27\mathfrak{q}^{4}}}}+\frac{1}{\sqrt[3]{2}}\sqrt[3]{7+27\mathfrak{q}^{2}-3\sqrt{3}\sqrt{3+14\mathfrak{q}^{2}+27\mathfrak{q}^{4}}}-3. \label{eq:7}
	\end{align}
Eq.  \eqref{eq:7} expresses the parameter $\beta$ expressed as a function of the tidal charge $\mathfrak{q}$, which plays the role of the charge wrapped by the event horizon. For both the values of $\beta$ in Eqs. (\ref{eq:6}, \ref{eq:7}), the surface dictated by the respective values of $r_{\beta}$ consists of a Killing horizon. 
Fig. \ref{dois} illustrates the parameter $\beta$ as a function of $\mathfrak{q}$. 

\begin{figure} [H]
	\centering
	\includegraphics[scale=0.55]{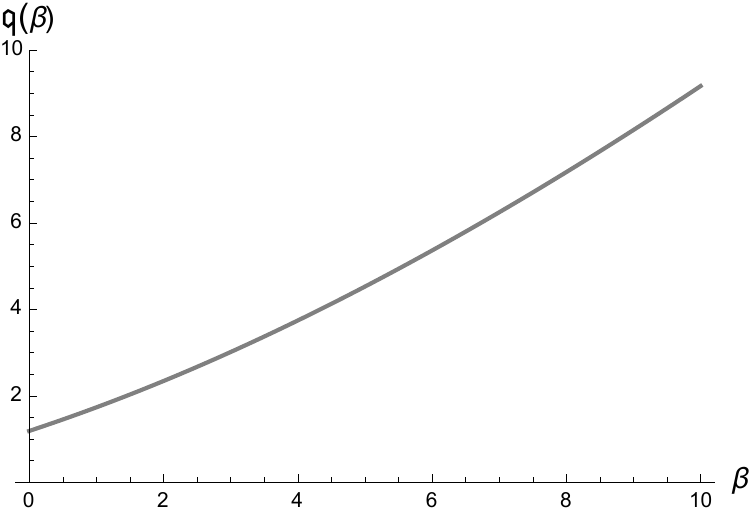}
	\caption{\clt{Plot of the parameter $\mathfrak{q}(\beta)$ as a function of $\beta$, in Eq.  \eqref{qbeta}.} }
		\label{doisq}
\end{figure}

\clt{As long as $\beta>-2$, as $\beta$ is the deformation parameter, the charge $\mathfrak{q}$ must be a function of $\beta$, and not the contrary. Therefore, using the implicit function theorem, one arrives at a more straightforward expression:} 
\clt{\beq\label{qbeta}
\mathfrak{q}(\beta)=\frac{1}{3 \sqrt{3}}\sqrt{\beta ^3+9 \beta ^2+33 \beta +38}.
\eeq}
Fig. \ref{doisq} illustrates the parameter $\beta$ as a function of $\mathfrak{q}$. 
\noindent
\noindent \clt{Figs. \ref{dois} and \ref{doisq} illustrates the saturation $\mathfrak{q}=\sqrt{3}$ at $\beta=1$.}

{\color{black}{
\par The electromagnetic potential has boundary components represented by the chemical potential and the charge
density, to wit, in the near-boundary expansion, 
\beq\label{lim1}
A(r)=\mu-\frac{\mathfrak{q}}{r} + \mathcal{O}\left(\frac1{r^2}\right).
\eeq 
To require a regular gauge connection, one must impose  $\lim_{r\to r_0}A(r)=0$ at the horizon, implying that 
$\mu=\frac{\mathfrak{q}}{r_0}$. Otherwise, calculating the holonomy of the electromagnetic potential around the (Euclidean) time loop would yield a non-vanishing value, when the loop collapses at the $r_0$ horizon, meaning a singular 
gauge connection.  
The planar \adr\, is described by two scales: the chemical potential $\mu = \lim_{r\to\infty}A(r)$ and the horizon radius $r_0$. 
For the \clt{\gbb},  the solution to the Maxwell equations $	\partial_{\mu}\left(\sqrt{-g}F^{\mu\nu}\right)=0$, for $F_{\mu\nu}=\partial_{[\mu}A_{\nu]}$, yields 
	\begin{equation} \label{eq:EMsol}
	\!\!\!\!\!A(r)\!=\!\alpha(\beta,r)\mathfrak{q}\!\left(\!2\sqrt{r\!-\!r_{0}}\sqrt{\beta\!+\!2}\sqrt{3r\!-\!r_{0}\left(\beta\!+\!2\right)}+r\left(\beta\!-\!1\right)\!\!\left[\pi\!-\!2\arctan\!\left(\!\frac{\sqrt{r\!-\!r_{0}(\beta\!+\!2)}}{\sqrt{3r\!-\!r_{0}\left(\beta\!+\!2\right)}}\right)\right]\right),
	\end{equation}
where
\beq
\alpha(\beta,r)=\frac{1}{2\sqrt{3}r_{0}\sqrt{\beta+2}\,r}.
\eeq
One can verify that
\begin{equation}\label{eq:EMsolLimit}
	\lim_{\beta\rightarrow1}A(r)=\mathfrak{q}\left(\frac{1}{r}-\frac{1}{r_{0}}\right)\text{ },
\end{equation}
corresponding to the AdS${}_4$--RN setup. 
The chemical potential, $\mu$, of the CFT${}_3$ in the \clt{\gbb}\,\!  boundary, can be obtained by the expansion of Eq. \eqref{eq:EMsol}, as
\begin{equation}
	\mu=-\frac{\mathfrak{q}}{6r_{0}}\left[6+\frac{\sqrt{3}\left(\beta-1\right)}{\sqrt{2+\beta}}\arctan\left(\sqrt{\frac{\beta+2}{3}}\right)\right]\text{ }.
\end{equation}

\par A range for the charge $\mathfrak{q}$ can be obtained, when  the shear viscosity-to-entropy density ratio is calculated for the \clt{\gbb}. The Kovtun--Son--Starinets (KSS) result, $\eta/s=1/4\pi$ \cite{kss}, will be shown to correspond to $\mathfrak{q}=\sqrt{3}$, by adding an off-diagonal gravitational perturbation $h_{xy}(t)$. The constitutive equation for the stress tensor with the dissipative term   is given by 
\begin{equation}
T^{\mu \nu} = (\rho + p) u^\mu u^\nu + P g^{\mu \nu} -  P^{\mu \sigma} P^{\nu \lambda} \left[\eta\left ( \nabla_{(\sigma} u_{\lambda)} - \frac{2}{3} g_{\sigma \lambda} \nabla_k u^k \right )
+\zeta g_{\sigma \lambda}  \nabla_k u^k\right],
\end{equation}
where $\nabla_\mu$ represents the covariant derivative with respect to the perturbed metric $g_{\mu \nu}$. The $P^{\mu \nu} := g^{\mu \nu} + u^\mu u^\nu$ is the projection tensor and  the shear, $\eta$, and the bulk, $\zeta$, viscosities carry dissipative effects.  Kubo's formula for the shear viscosity $\eta$ can be derived by coupling fictitious gravity to the fluid and then determining the response of $T^{\mu \nu}$ under small gravitational perturbations. 
The covariant derivative of the velocity field reads  $\nabla_\mu u_\nu = \Gamma^t_{\mu \nu}$, assuming $u_k=u_k(t)$. Calculating the response in $T^{xy}$ yields 
\begin{equation}
\delta \left \langle T^{xy} \right \rangle = -2 \eta \Gamma^t_{xy} = -\eta \partial_t h_{xy} \ ,
\end{equation} 
whose Fourier transform is given by $
\delta \left \langle T^{xy} (\omega) \right \rangle = i \omega \eta h_{xy}$. Comparing this equation to $\delta \left \langle T^{xy} \right \rangle = - G_R^{xy, xy} h_{xy}$ yields the Kubo's formula for the shear viscosity  \cite{kss,natsuume} 
\begin{equation}
\eta = -\lim_{\substack{{\omega}\rightarrow 0\\
 k\to0}} \frac{1}{\omega} \Im G_R^{xy, xy} (\omega, \vec{k}).
\end{equation}

AdS/CFT duality states that the partition functions of both the gauge and the gravitational theories are equivalent objects, by the so-called   GKPW relation \cite{malda,gkp1,gkp2},
\begin{equation}
\left \langle e^{i \int \varphi^{\scalebox{.65}{(0)}} \mathcal{O}}\right \rangle = e^{i \bar{S}[\varphi^{\scalebox{.65}{(0)}}]}.
\label{eq:gkp_witten}
\end{equation}
Here $\bar{S}$ denotes the on-shell action, whereas the scalar field in the bulk theory of gravity is denoted by $\varphi$. With the definition $u=r_{0}/r$, one can denote $\varphi^{\scalebox{.65}{(0)}} = \lim_{u\to0}\varphi$. 
The gravitational theory is the Einstein--Hilbert action and a term describing matter, 
\begin{equation}\label{eq:fullAction}
	S=\frac{1}{16\pi G_4}\int \sqrt{-g}\left(R-2\Lambda_4\right)\,d^{4}x+S_{\scalebox{.65}{\textsc{mat}}},
\end{equation} where denoting by a prime the differentiation with respect to $u$, 
\begin{eqnarray} \label{eq:assympAction}
	S_{\scalebox{.65}{\textsc{mat}}}=-\frac{1}{2}\int \sqrt{-g}\nabla_{\mu}\varphi\nabla^{\mu}\varphi\,d^{4}x = -\frac{1}{2}\int \frac{r_{0}^{3}}{u^{2}}\varphi^{\prime\,2}\,dx^{4}.
\end{eqnarray} The last equality in Eq. (\ref{eq:assympAction}) comes from requiring the scalar field $\varphi=\varphi\left(u\right)$ to be static. 
Integrating by parts and assuming the scalar field fades away as it approaches its null value at the horizon yields the equation of motion $\frac12\left(\varphi^{\prime}/u^{3}\right)^{\prime}\sim0$, whose asymptotic behavior is given by 
\begin{equation} \label{eq:scalarSol}
	\varphi\sim\varphi^{\left(0\right)}\left(1+\varphi^{\left(1\right)}u^{3}\right) \ .
\end{equation}
Therefore, the action \eqref{eq:assympAction} reduces to a surface term on the AdS boundary. Substituting it in Eq. \eqref{eq:scalarSol}, the on-shell action reads 
\begin{equation} \label{eq:actionOnshell}
	\bar{S}\left[\varphi^{\left(0\right)}\right]=\frac{3}{2}\int r_{0}^{3}\left(\varphi^{\left(0\right)}\right)^{2}\varphi^{\left(1\right)}\,d^{3}x,
\end{equation}
whereas the one-point function has the form
\begin{equation} \label{scalarOnepoint}
	\left\langle \mathcal{O}\right\rangle _{S}=3r_{0}^{3}\varphi^{\left(1\right)}\varphi^{\left(0\right)} \ .
\end{equation}
Compared to the linear response relation  yields the Green function $
\lim_{k\to0}G_R^{\mathcal{O}, \mathcal{O}} (k) = -3 r_0^3 \varphi^{\scalebox{.65}{(1)}}$.

The $\eta/s$ ratio is used to delimit the parameter $\beta=\beta(\mathfrak{q})$ of generalized AdS$_4$ black branes. 
For it, a bulk small gravitational perturbation $h_{xy}$ can be considered, such that
$
ds^2 = ds^2_{0} + 2h_{xy} dx dy$, 
where $ds^{2}_{0}$ is given by Eq. \eqref{rngen}. The response of the energy-momentum tensor on the boundary wits 
\begin{equation}
\delta \left \langle T^{xy} \right \rangle  = i \omega \eta h_{xy},
\label{eq:response_1_2}
\end{equation}
for $h_{xy}^{\scalebox{.65}{(0)}}$ being the perturbation to the boundary theory, which is asymptotically related to $h_{xy}$ by
the result $g^{xx}h_{xy} \sim h_{xy}^{\scalebox{.65}{(0)}} \left ( 1 + h_{xy}^{\scalebox{.65}{(1)}} u^3 \right)$. 
 Eq.  \eqref{scalarOnepoint} yields 
\begin{equation}
\delta \left \langle T^{xy} \right \rangle = \frac{r_0^3}{16 \pi}3 h_{xy}^{\scalebox{.65}{(1)}}h_{xy}.
\label{eq:response_2_2}
\end{equation}
Comparing Eqs.  \eqref{eq:response_1_2} and Eq. \eqref{eq:response_2_2} implies that $
i \omega \eta = \frac{3r_0^3}{16\pi G_4} h_{xy}^{\scalebox{.65}{(1)}}.$ Taking into account that the entropy density associated with Eq. \eqref{rngen} reads $s = r_{\beta}^2/4$, implies  
\begin{equation}
\begin{gathered}
\frac{\eta}{s} = -\frac{3ir_0}{4\pi\left[1+\frac{2}{9}\left(\beta-1\right)\right]^{2}} \frac{h_{xy}^{\scalebox{.65}{(1)}}}{\omega} \ .
\label{eq:eta_s_geral_2}
\end{gathered}
\end{equation}
\par One now must find $h_{xy}^{\scalebox{.65}{(1)}}$, solving the equation of motion for the 4D massless scalar perturbation $g^{xx}h_{xy}$. Taking $\varphi(u) = \phi(u) e^{-i\omega t}$, the perturbation reads 
\begin{equation}
\phi = \phi^{\scalebox{.65}{(0)}} \left[1 - i \frac{\omega}{r_0} \frac{\mathfrak{q}^2-3}{|\mathfrak{q}^2-3|} \bigintsss {u^2}\sqrt{{\left\{ \frac{1-u}{1-u\left[1+\frac{1}{3}\left(\beta-1\right)\right]}\right\}}}\, d u \right],
\end{equation}
and 
\begin{equation}
h_{xy}^{\scalebox{.65}{(1)}} = -\frac{i\omega}{3r_0}  \frac{\mathfrak{q}^2-3}{|\mathfrak{q}^2-3|} \ .\label{abo}
\end{equation}
\noindent When Eq. (\ref{abo}) is superseded into Eq.  \eqref{eq:eta_s_geral_2}, it yields
	\begin{equation} \label{eta-sRatio}
		\frac{\eta}{s}=\color{black}{9}{\left(1-\frac{2\sqrt[3]{2}}{\alpha_{\mathfrak{q}}}+\frac{\alpha_{\mathfrak{q}}}{\sqrt[3]{2}}\right)^{\!\!-2}} \frac{({3-\mathfrak{q}^{2}})}{\left|3-\mathfrak{q}^{2}\right|}\ ,
	\end{equation}
	for $\color{black}\alpha_\mathfrak{q}=\sqrt[3]{-7-27\mathfrak{q}^{2}+3\sqrt{9+42\mathfrak{q}^{2}+81\mathfrak{q}^{4}}}$, 
where the $\beta=\beta(\mathfrak{q})$ parameter was written as given by Eq.  \eqref{eq:7}. 
\begin{figure} [H]
	\centering
	\includegraphics[scale=0.65]{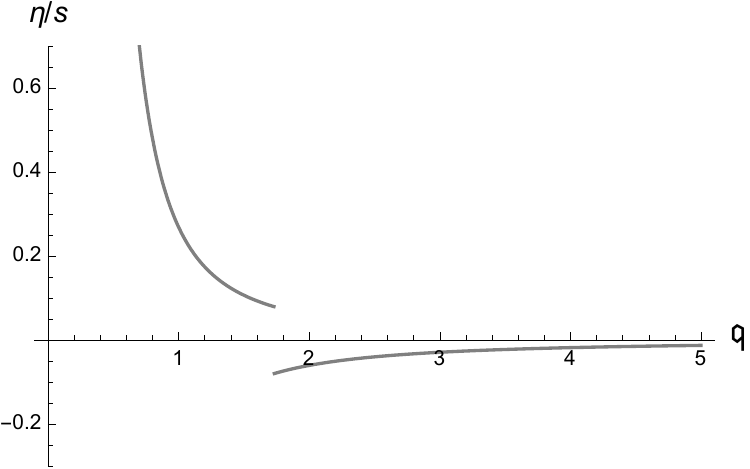}
	\caption[short text]{Plot of $\eta/s$ as a function of $\mathfrak{q}$, as in Eq.  \eqref{eta-sRatio}. One can realize that the $\eta/s$ ratio changes sign at $\mathfrak{q}=\sqrt{3}$.}	\label{eta-s-plot1}
\end{figure}
\noindent Fig.  \ref{eta-s-plot1} transport at the extremal horizon as well (via the membrane paradigm), shows that the bound $
0 < \mathfrak{q} < \sqrt{3}$ 
yields a positive value of the ratio $\eta/s$, which assumes the saturated KSS value $\eta/s = 1/4\pi$ when the tidal charge saturates, namely, $\mathfrak{q}= \sqrt{3}$. \clt{The thermodynamically correct range $0 < \mathfrak{q} < \sqrt{3}$ implies $-2< \beta <1$  for the \clt{\gbb}, yielding the event horizon $r_\beta$ to be smaller than $r_0$}, being an inner horizon. It indeed complies with the interpretation of $r_0$ as the effective outermost of these event horizons. Therefore, in the next section, the term near-horizon corresponds to $u=r_0/r\to1$.

Similarly, the bulk viscosity can be calculated by the Kubo's formula
\beq{\zeta}
=\lim_{\substack{{\omega}\rightarrow 0\\
 k\to0}} \frac{1}{\omega}\Im G_R^{PP}(\omega, \vec{k}),\eeq
 where 
 \beq
 G_R^{PP}(\omega, \vec{k}) &=& \frac{k_ik_jk_mk_n}{k^4}\left[G_R^{ij,mn}(\omega, \vec{k})+\frac13 \delta_{ab}T^{ab}\left(\delta^{i(m}\delta^{j|n)}-\delta^{ij}\delta^{mn} \right)\right] +\frac13 \delta_{ij}T^{ij} -\frac43 G_R^{xy,xy}(\omega, \vec{k})\nonumber
 \eeq
is the response function to longitudinal fluctuations \cite{Czajka:2017bod}.  
The bulk viscosity-to-entropy density ratio reads, for the \clt{\gbb}, 
	\begin{equation} \label{zeta-sRatio}
		\frac{\zeta}{s}=\frac{3}{16}\left(1-\sqrt[3]{2}{\gamma_{\mathfrak{q}}}+\frac{1}{\sqrt[3]{2} \gamma_{\mathfrak{q}}}\right) (3-\mathfrak{q}^{2})^2\ ,
	\end{equation}
	where $\color{black}\gamma_\mathfrak{q}=\sqrt[3]{4+14\mathfrak{q}^{2}+\sqrt{4+14\mathfrak{q}^{2}+46\mathfrak{q}^{4}+12\mathfrak{q}^{6} }}$, is plotted in Fig. \ref{zeta-s-plot1}, in the allowed range $0 < \mathfrak{q} < \sqrt{3}$ where $\eta/s>0$.
\begin{figure} [H]
	\centering
	\includegraphics[scale=0.65]{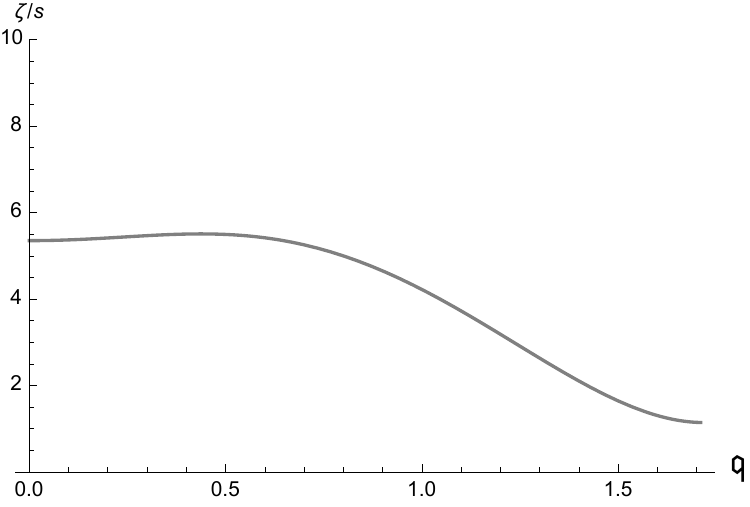}
	\caption[short text]{Plot of $\zeta/s$ as a function of $\mathfrak{q}$, as in Eq.  \eqref{zeta-sRatio}. }	\label{zeta-s-plot1}
\end{figure}

\par The temperature of the \clt{\gbb}\,\! at the Killing horizon $r_{\beta}$ can be evaluated by  \cite{Ferreira-Martins:2019wym}
\begin{equation} \label{eq:8}
	\upkappa^{2}=\lim_{r\rightarrow r_{\beta}}\left[-\frac{\left(\upxi^{\mu}\nabla_{\mu}\upxi^{\lambda}\right)\left(\upxi^{\nu}\nabla_{\nu}\upxi_{\lambda}\right)}{\upxi^{\rho}\upxi_{\rho}}\right].
\end{equation} In this way $\upkappa$ is the surface gravity at the horizon \cite{wald}, which is related to the temperature of \clt{\gbbs}\,\!  by 
\begin{equation} \label{eq:9}
	T=\frac{\upkappa}{2\pi}\ .
\end{equation}
Evaluating \eqref{eq:8}  at the horizon $r=r_\beta$ yields 
\begin{align}
\begin{aligned}
\upkappa^{2}&=\lim_{r\to r_{\beta}}\frac{\left[2r^{4}+\left(1+\mathfrak{q}^{2}\right)r_{0}^{3} r-2\mathfrak{q}^{2}r_{0}^{4}\right]^{2}\left[3r+\left(\beta+2\right)r_{0}\right]}{12r^{4}\left(r-r_{0}\right)^{2}\left(r^{3}+r^{2}r_{0}+rr_{0}^{2}-\mathfrak{q}^{2}r_{0}^{3}\right)}=0\label{tempads}
\end{aligned}
\end{align}
Since Eq. (\ref{tempads}) is independent of $\beta$, it holds for the entire range $-2< \beta < 1$.
A zero-temperature black brane is expected, when extremal \clt{\gbbs}\,\!           
 are regarded.  In the extremal case where the electromagnetic field carries all its energy, 
the extremal AdS$_{4}$-RN black brane corresponds to the zero-temperature state of the boundary \cite{Davison:2013txa}. The metric \eqref{rnads4} with coefficients \eqref{rnads41} define the extremal AdS$_{4}$-RN black brane with the planar horizon, with  mass coincident with its underlying electromagnetic energy. Therefore, it no longer sheds energy, owing to charge conservation. Consequently, the Hawking temperature of AdS$_{4}$-RN black branes vanishes as $r\to r_\beta$ since they are forbidden to radiate due to charge conservation. An analogous process occurs in the family of extremal AdS$_4$ generalized black branes described by the metric \eqref{rngen}, with the metric temporal coefficient $f(r)$ as in Eq. \eqref{rnads41}, but with metric radial coefficient \eqref{eq:1}.
The  associated Hawking temperature equals zero, as $r\to r_\beta$. The extremal AdS$_{4}$-RN black branes  have their whole mass due to their conserved electric charge. Hence, further reducing the mass becomes impossible, and energy conservation prevents  radiation emission. The horizon has a lower limit, having macroscopic size, implying that the extremal AdS$_{4}$-RN black brane does carry a finite Bekenstein--Hawking entropy. Therefore, this feature does allow the deformation parameter $\beta$ to attain the value $\beta = -2$, since for it $r_\beta = 0$ and the black brane charge vanishes, $\mathfrak{q}=0$, according to Eq. (\ref{eq:4}) and Fig. \ref{dois}. Also, near $\beta = -2$, the horizon $r_\beta$ would not be macroscopic. Therefore, one can stipulate a reasonable range 
\beq
-1.95\lesssim \beta < 1,\label{195}
\eeq
compatible with a macroscopic horizon. In order to study systems in condensed matter, the infrared (IR) limit has a deep interest, corresponding to the near-horizon limit of AdS$_4$ generalized black branes.  Like the standard AdS$_{4}$-RN, the near horizon structure is AdS${}_2\times \mathbb{R}^2$ in the range \eqref{195}.

Now, one can better clarify aspects of the $\beta$ parameter.
AdS/CFT relates the electric part of the Weyl tensor, $\mathcal{E}_{\mu\nu}$,  representing the propagation of classical gravitational waves in the AdS$_5$ bulk, to the expectation value $\langle T_{\mu\nu}\rangle$ of the renormalized energy-momentum tensor of conformal fields on the AdS$_4$. The large-$N$ limit expansion in  CFT requires $N\sim 1/(T_4\ell_p)^2\gg1$. In the original Randall--Sundrum braneworld models, the Planck length, $\ell_p$ (for $8\pi G_4 = \ell_p^2$, where $G_4$ is the 4D Newton's constant), is related to the AdS$_5$ Planck length $\ell_5$ by $\ell^2_p = T_4 \ell^3_5$ \cite{ssm2}, where $T_4$ is the \clt{\gbb}\,\! tension  \cite{Shiromizu:2001jm,Kanno:2002iaa}. In addition, embedding AdS$_4$ in AdS$_5$ introduces a normalizable 4D graviton and an ultraviolet (UV) cut-off in the CFT, which is proportional to $T_4^{-1}$. In the AdS/CFT setup, 
$\mathcal{E}^\nu_{\;\mu}\sim\ell_p^2 \langle T_{\;\mu}^{\nu}\rangle$. 
Since the electric part of the Weyl tensor is traceless, it implies that $\langle T_{\;\mu}^{\mu}\rangle=0$, which is valid as long as the conformal
symmetry is not anomalous.  One can use the GKPW relation to consider the partition function of the dual theory at the boundary. On the other hand, statistical-mechanical arguments state that the partition function is proportional to the free energy, $F$, which yields the energy density and the pressure. For  \clt{\gbbs}, a conformal anomaly encodes quantum corrections induced by $\beta\neq 1$, to wit
\beq
\langle T_{\;\mu}^{\mu}\rangle=\left[\left(4 \mathfrak{q}^2 -3 \left(\mathfrak{q}^2+1\right) \right) \left(\mathfrak{q}^2 -3\right)^2 ((3 \beta +2)
   -3)^2\right]\left(\frac{3+\beta}{4+3\beta}\right)^{2}.\eeq
It complies with the fact that the vanishing of $ \langle T_{\;\mu}^{\mu} \rangle$ implies the UV cutoff must be smaller than any physical length scale involved \cite{Kuntz:2019omq,Kuntz:2022kcw}. Also, $\langle T_{\;\mu}^{\mu}\rangle\neq0$ signs the presence of an intrinsic
 length. Otherwise, the CFT could be affected by that scale.
For the \clt{\gbb}, the horizon $r_0$ is a natural length scale, and one therefore expects that only wave modes in the CFT having wavelength $\ell\ll r_0$, but still larger than $T_4^{-1}$, can freely propagate \cite{Meert:2020sqv,Meert:2021khi}.  

For the  \clt{\gbb}, the holographic computation of the 
 Weyl anomaly can be implemented \cite{Henningson:1998gx,Kanno:2002iaa}. Denoting $a$ and $c$ central charges of the conformal gauge theory (see Eq. (24) of Ref. \cite{Cremonini:2011iq}), 
 \beq\label{wa}
\!\!\!\!\!\!\!\! \langle T^\mu_{\;\,\mu}\rangle_{\scalebox{.65}{\textsc{CFT}}} &=& \frac{c}{16\pi^2}\left(R_{\mu\nu\rho\sigma}R^{\mu\nu\rho\sigma}-2R_{\mu\nu}R^{\mu\nu}+\frac13 R^2\right) -\frac{a}{16\pi^2}\left(R_{\mu\nu\rho\sigma}R^{\mu\nu\rho\sigma}\!-\!4R_{\mu\nu}R^{\mu\nu}\!+\!R^2\right),
 \eeq where the terms in parentheses are, respectively, the Euler density and the square of the Weyl curvature \cite{Casadio:2003jc}.  
The \clt{\gbb}\,\! at the boundary $u\to0$, can have the square of the Weyl curvature to be Taylor-expanded as 
\beq\label{430}
\!\!\!\!\!\!\!&&N^2\left[\frac{40}{3}+\frac{64}{9} (3 \beta -1) u+\frac{2}{27} \left(315 \beta ^2+6 \beta -37\right) u^2+\frac{8}{81} u^3 \left[27 \beta  (\beta  (7 \beta +4)-1)+189
   \mathfrak{q}^2+179\right]\right.\nonumber\\&&\left.\qquad+\frac{1}{243} u^4 \left[(3 \beta -1) (36 \beta  (3 \beta  (5 \beta +9)+16)+1219)+27 (123 \beta -137) \mathfrak{q}^2\right]\right]+\mathcal{O}\left(u^5\right),
\eeq
and  the Euler density reading 
\beq\label{120}
&&N^2\left[120+64 (3 \beta -1) u+24 \left(9 \beta ^2-1\right) u^2+u^3 \left(32 \beta  \left(6 \beta ^2+3 \beta -1\right)+168 \mathfrak{q}^2+\frac{1448}{9}\right)\right.\nonumber\\&&\left.\qquad +u^4 \left(\frac{2}{27} (3
   \beta -1) (36 \beta  (3 \beta  (5 \beta +7)+8)+853)+2 (93 \beta -79) \mathfrak{q}^2\right)\right]+\mathcal{O}\left(u^5\right),
\eeq where $N^2 = \pi L^3/2G_4$.
Therefore, the CFT boundary $u\to0$ limit implies that  
\beq
 \langle T^\mu_{\;\,\mu}\rangle_{\scalebox{.65}{\textsc{CFT}}} =\frac{400 N^2}3,\eeq
having an analogous result to the \adr. In the next section, we will study the condensed matter state dual to extremal AdS$_4$ generalized black branes. 

\section{Holographic superconductor and AdS$_4$ generalized black branes}
\label{hs1ads}
Now, one can turn into the holographic superconductor, with the AdS${}_4$ generalized  black brane background. As reported by the Ginzburg--Landau theory, superconductivity relies on the (spontaneous) U(1)\;symmetry breaking leading to a superfluid states, as long as the U(1) broken symmetry is global. Nevertheless, the superfluid state can be promoted to a superconductor by the weak coupling of a U(1) gauge field in the boundary theory, with a conserved U(1) current. The gauge symmetry in the dual bulk is broken, by a subsequent formation of a Higgs condensate in the bulk \cite{Hartnoll:2008vx}. This setup enables to describe superconductivity in the bulk, by adding a charged complex scalar field$, \upphi$, emulating the Higgs field in Ginzburg--Landau theory, minimally coupled to gravity, and a Maxwell Yang--Mills field as well. The 4-dimensional $s$-wave holographic superconductor (HS) in the AdS$_4$ bulk embedded in an AdS$_5$ space can be described by
\begin{subequations}
\begin{align}
S_{\scalebox{.65}{\textsc{bulk}}} 
&= \int  \left(R-2\Lambda_4\right)\,\sqrt{-g}\,d^4x+S_{\scalebox{.65}{\textsc{HS}}} ~, \\
S_{\scalebox{.65}{\textsc{HS}}}  &= -\frac{1}{g_{\scalebox{.65}{\textsc{YM}}}^2} \int  \sqrt{-g} \left( \frac{1}{4}F_{\mu\nu}F^{\mu\nu} + D^\mu\upphi D_\mu\upphi^*+m^2 |\upphi|^2 \right)\,d^4x,
\label{eq:10}
\end{align}
\end{subequations}
where $g_{\scalebox{.65}{\textsc{YM}}}$ is an effective coupling. Here $F_{\mu\nu}=\partial_{[\mu}A_{\nu]}$ is the U(1)\;gauge field strength associated with the U(1)\;gauge field $A_\mu$ that describes finite charge density. The Maxwell term lies in the part $S_{\scalebox{.65}{\textsc{HS}}}$ of the action in Eq. (\ref{eq:10})  and it does not contribute to the computation of the shear viscosity, $\eta$. 
The bulk scalar field $\upphi$ is the dual object to the order parameter, whose VEV parametrizes the broken U(1)\;global symmetry in the boundary.  It is worth mentioning that the potential $V(\upphi) = m^2 |\upphi|^2$ in (\ref{eq:10}) refers to the so-called minimal holographic superconductor, where one disregards self-interaction (and higher-order integer powers) of the Higgs scalar. The  \clt{\gbb}\,, presented in the previous section, will be used to study the holographic superconductor for the probe limit $g_{\scalebox{.65}{\textsc{YM}}}\gg 1$ yielding no backreaction of the scalar and gauge fields onto the \clt{\gbb}\,\! geometry. Expressing the extrinsic curvature, coming from the AdS$_4$ space embedded in AdS$_5$, by  \cite{maartens}
\beq\label{ec}
K_{\mu\nu}=-\frac{1}{2}\left[T_{\mu\nu}+\frac13\left(T_4-T^\alpha_{\;\alpha}\right)g_{\mu\nu}\right],
\eeq
where $T_4$ is the \clt{\gbb}\,\! intrinsic tension yields the electric part of the Weyl tensor 
to be written as 
\beq\label{wt}
E_{\mu\nu}=-\frac{\Lambda}{6}g_{\mu\nu}-\partial_w K_{\mu\nu}+K_\mu^{\;\rho}K_{\rho\nu},
\eeq
where $w$ denotes the Gaussian coordinates along the AdS$_5$ bulk, as previously mentioned.  Therefore, the bulk equations of motion for the matter fields are obtained from the action \eqref{eq:10} and expressed as
\begin{subequations}
\begin{align}
D_\mu D^\mu \upphi-m^2\upphi&=0~,\label{eqq1} \\
\nabla^\mu F_{\mu\nu} - 2\Im(\upphi^\dagger D_\nu\upphi)&=0,\label{eqq2}
\end{align}
\end{subequations}
with covariant derivative $D_\mu\upphi = \partial_\mu\upphi + iqA_\mu\upphi$, where $q$ represents the scalar field charge. 
Assuming a central scalar field $\upphi=\upphi(r)$ and an electrostatic potential $A_t\neq0$, Eq. (\ref{eqq2}) yields 
\beq\label{atat}
 -m^2\upphi +q^2 f(r) (A^t) ^2\upphi =0,
\eeq
for $f(r)$ being the time component of the \clt{\gbb}\,\! metric, given by Eq. (\ref{rnads41}). 
It suggests that the electrostatic potential plays the role of a negative mass-squared, characterizing the charged scalar field. If it gets large enough, it can stir up instabilities. 
For $\beta \approx 1$,  the AdS$_4$ generalized black brane (\ref{rngen}) with metric terms (\ref{rnads41}, \ref{eq:1}) can be described by a domain wall playing the role of a tunneling barrier, going from a pure AdS$_4$ bulk, in the UV regime, and the near-horizon AdS$_2 \times \mathbb{R}^2$ topology, in the IR limit \cite{Hartnoll:2009sz}. For any value of $\beta$, in the pure  AdS$_4$ bulk case, taking into account the electromagnetic potential in Eq. \eqref{eq:EMsol}, the term $f(r) (A^t)^2$ in Eq. (\ref{atat}) contributes as
\beq\label{pote2}
&&\frac{1}{12 (\beta +2) R^2}\left\{q^2(\beta -1) \left[\frac{5}{{4(\beta +2})^{3/2} \sqrt{R} (\beta  R+2 R-1)}-\frac{\beta -1}{2}\right.\right.\nonumber\\&&\left.\left.+\left((\beta -1) \left(2 {\rm arctanh}\left(\frac{{1}}{{\sqrt{(\beta +2) R}}}\right)+\pi \right)-(\beta +2)\right)^2\right]\right\}\times\frac{1}{r^2}.
\eeq
and is subleading to the mass term in the UV limit. The limit $\beta\to -2$  makes this term diverge, and a secure neighborhood of the value $\beta = -2$ should be avoided, again corroborating with the range (\ref{195}). As the potential (\ref{pote2}) goes as $\propto 1/r^2$ as $r\to\infty$, the quantum theory is stable and the Breitenlohner--Freedman (BF) bound for the AdS$_4$ generalized black brane reads\footnote{Remember that here we assumed $L=1$ and, correspondingly, $m^2L^2\geq -\frac94$, for the most general case.}
\beq\label{bf}
m^2\gtrsim -\frac94,
\eeq
analogously to the AdS$_4$-RN black brane case, irrespectively of the value of $\beta$.
The vacuum is stable if the BF bound (\ref{bf}) holds. 
For analyzing the IR limit, AdS$_2 \times \mathbb{R}^2$,  the term $f(r) (A^t)^2$ in Eq. (\ref{atat}) is a constant. For values of $\beta\approx 1$,  a second stability criterium exists for the extremal  AdS$_4$ generalized black brane in the IR geometry, namely
\beq
m^2\geq-\frac32+\frac{q^2g_{\scalebox{.65}{\textsc{YM}}}^2L^2}{8\pi G_4}.
\eeq
Therefore, the charged scalar field coupled to AdS$_4$ generalized black brane geometry can be unstable in the IR limit and stable in the UV regime. When the BF bound is violated, in the IR limit, spontaneous pair production sets in. The corresponding scalar field quanta can form a macroscopic ground state condensate, equivalent to the Green's function of the scalar field having its pole transferred to the upper half plane \cite{Faulkner:2009wj}. Near the charged extremal  AdS$_4$ generalized black brane horizon $r_\beta$, the electric field is strong enough to make the local vacuum discharge, emulating superradiance. The positively charged quanta escape and constitute AdS$_4$ generalized black brane hair. The AdS$_4$ geometry plays the role of a near-infinity gravitational potential barrier balancing the repulsive electromagnetic force, developing a charged scalar atmosphere around the AdS$_4$ generalized black brane. The charge is carried to the atmosphere, decreasing the near-horizon electric field intensity, as long as the pair production phenomenon does not come to an end.  In this way, the charged scalar field attains a finite value of amplitude, which is maximal at the near-horizon limit and corresponds to scalar hair. 

Let us suppose that $m^2<0$, above the BF bound (\ref{bf}). 
In the decoupling limit $8\pi G_4g_{\scalebox{.65}{\textsc{YM}}}^2L^2 \gg 1$ of weakly-coupled gravity, the scalar and gauge sectors become decoupled from gravity and do not contribute to the AdS$_4$ curvature. Therefore, all the analytical and numerical methods can be used in the fixed background of the \clt{\gbb}\,\! with metric \eqref{rngen}, with  coefficients (\ref{rnads41}, \ref{eq:1}). Using the notation  
\begin{subequations}
\begin{align}
\upvarphi&=\frac{\upphi}{u},\\
V(u)&=\frac{1}{u^2}\left(m^2+2\sqrt{f(u)n(u)}-\frac{\left(f'(u)n(u)+n'(u)f(u)\right)u}{2\sqrt{f(u)n(u)}}\right),
\end{align}
\end{subequations}
in the $A_u = 0$ gauge 
the static bulk equations (\ref{eqq1}, \ref{eqq2}) read
\begin{subequations}
\label{eq:eom_bulk}
\beq
\left[-\sqrt{f(u)n(u)}\partial_u^2 - \partial_x^2-\partial_y^2 + 2|\upvarphi|^2\right]A_t&=&0~, \\
\left[ -\partial_u(\sqrt{f(u)n(u)}\partial_u) - \partial_x^2-\partial_y^2 +2|\upvarphi|^2 \right] A_i -2\Im(\upvarphi^\dagger \partial_i\upvarphi) + \partial_i\left(\delta^{jk}\partial_j A_k\right)&=&0~, \\
 \left[ -\partial_u(\sqrt{f(u)n(u)}\partial_u)  - \frac{A_t^2}{\sqrt{f(u)n(u)}} - \delta^{ij}D_iD_j + V(u)\right] \upvarphi&=&0~, \\
\partial_u\left[\left(\delta^{ij}\partial_i A_j\right)\right] - 2 \Im(\upvarphi^\dagger\partial_u\upvarphi)&=&0~,
\eeq
\end{subequations}
The CFT two-point correlation function for scalar operators with scaling dimension can be expressed in terms of the ratio of the coefficients
$\upphi^{(\pm)}$ of the leading and sub-leading asymptotes of the corresponding AdS massive scalar waves. In
the near-boundary region one can express $\langle \mathcal{O}(-k)\mathcal{O}(k)\rangle = \upphi^{(+)}/\upphi^{(-)}$, for the bare (UV) scalar operator $\mathcal{O}(x)$ of the CFT with conformal scaling dimension $\Delta$. Therefore, both the electromagnetic potential and the scalar field wit  
\begin{subequations}
\label{eq:dictionary}
\begin{align}
A_\mu &\approx {\mathsf{A}}_\mu + A_\mu^{(+)} u~, \label{am1}\\
\upphi &\approx \upphi^{(-)} u^{\Delta_-} + \upphi^{(+)} u^{\Delta_+}~, \label{am2}
%
\end{align}
\end{subequations}
with scaling dimension $\Delta_\pm = \frac{3}{2}\pm\sqrt{\frac{9}{4}+m^2}$, 
set by the mass of the AdS scalar wave. 
In Eq. (\ref{am1}) ${\mathsf{A}}_t=\mu$ is the chemical potential, and $A_t^{(+)}$ represents the charge density $\langle\rho\rangle$, whereas $\mathsf{A}_i$ is the vector part of the electromagnetic potential and $A_i^{(+)}$ represents the current density $\langle{\mathsf{J}}_i\rangle$. In Eq. (\ref{am2}) the term 
$\upphi^{(+)}$ represents the order parameter $\langle{\mathcal{O}}\rangle$, whereas $\upphi^{(-)}$ is the external source for the order parameter \cite{Davison:2013txa}. 
According to the standard AdS/CFT,
\begin{subequations}
\begin{align}
\langle {\mathsf{J}}_\mu\rangle &= \frac{1}{g_{\scalebox{.65}{\textsc{YM}}}}\lim_{u\to0} F_{u\mu}.
%
\end{align}
\end{subequations}
The limit $g_{\scalebox{.65}{\textsc{YM}}}^2\to1$ is assumed in what follows, for the sake of simplicity.

\subsection{Small values of the magnetic field}
\label{3as}
The limit $\beta \to 1$ in Eq. (\ref{eq:6}) recovers the AdS${}_4$ black brane solution, 
matching the well-known results in Ref. \cite{Natsuume:2022kic}. One can argue whether magnetic fields can percolate into the holographic superconductor. For a small magnetic field, the results for the \clt{\gbb}\,\! are analogous to the standard AdS${}_4$-RN black brane solution. 
\clt{One can consider the system to be at zero temperature, exploring the $T = 0$ part of the phase diagram, since the focus hereon will comprise the magnetic field, the critical values of the magnetic field, $H_{c1}$ and $H_{c2}$, the Meissner effect, and creation of vortices.
The $T=0$ temperature is consistent with a uniform scalar condensate $\upvarphi_0 =\upvarphi_0(u)$ satisfying the equations of motion}. 
One can then apply a magnetic field perturbatively, considering $A_y=A_y(x,u)$ and $B_z=\partial_x A_y$. 
Therefore, taking into account the \clt{\gbb}\,\! metric coefficients  (\ref{rnads41}, \ref{eq:1}), the bulk Maxwell equations can be expressed as
\begin{align}
\left[ -\partial_u(\sqrt{f(u)n(u)}\partial_u) +k^2 + 2|\upvarphi_0|^2 \right] \mathring{A}_y = 0,
\label{bme}
\end{align}
where \beq
\mathring{A}_y(k,u)=\frac1{2\pi}\int_\mathbb{R} e^{-ikx}{A}_y(x,u)\,dx.\eeq 
For $r\gg r_\beta$, the solution of Eq.  \eqref{bme} reads
\beq
\mathring{A}_y(k,u)\!\!&\!=\!&\! c_1 U\left(-\frac{9 \beta ^2-6 \beta-4 k^2  -8
   |\upvarphi_0|^2 -15}{(\beta +1) (3 \beta -1)},\frac{2 (3 \beta -7)}{3 \beta -1}+1,\frac{3}{2} u
   (\beta +1)-\frac{9 (\beta +1)}{3 \beta -1}\right)\nonumber\\
   &&\!\!\times \exp\! \left(\frac{-9\beta}{2 (1\!-\!3 \beta )^2}\left[\left(3 \beta ^2\!+\!2
   \beta \!-\!1\right) u\!+\!(5\!-\!3 \beta ) \log ((1\!-\!3 \beta)  u\!+\!6)+(3 \beta \!-\!5) \log (3 \beta 
   u\!-\!u\!+\!6)\right]\right)\nonumber\\&+&c_2 L_{\frac{-4 k^2+9
   \beta ^2-6 \beta -8 |\upvarphi_0|^2 -15}{(\beta +1) (3 \beta -1)}}^{\frac{2 (3 \beta -7)}{3
   \beta \!-\!1}}\left(\frac{3}{2} u (\beta \!+\!1)-\frac{9 (\beta +1)}{3 \beta
   -1}\right)\nonumber\\
   &&\!\!\times \exp\! \left(\frac{3}{2 (1\!-\!3 \beta )}\left[\left(3 \beta ^2\!+\!2 \beta \!-\!1\right) u\!+\!(5\!-\!3 \beta ) \log ((1\!-\!3 \beta)  u\!+\!6)\!+\!(3
   \beta \!-\!5) \log (3 \beta  u\!-\!u\!+\!6)\right]\!\right),\nonumber\\\eeq
where $U[a,b,z]$ denotes the confluent hypergeometric function and $L^n_m[z]$ is the generalized Laguerre polynomial, with $c_1$ and $c_2$ integration constants.

One of the theoretical explanations of the Meissner effect, which is the physical property that makes the condensate repel magnetic flux, comes from the London equation, which states that the magnetic field strength decays inside the superconductor. The Meissner effect distinguishes superconductors from superfluids. It happens in superconductors with gauged symmetry. Under the usual Dirichlet boundary condition, no Meissner effect sets in: the magnetic field percolates the holographic superconductor. 
Eq. \eqref{bme} can be integrated as
\begin{align}\label{fo}
\sqrt{f(u)n(u)} \partial_u \mathring{A}_y =-\int_u^1  \left(k^2+2|\upvarphi_0|^2\right) \mathring{A}_y(u')\,du'\,.
\end{align}
Eq. (\ref{fo}) can be then integrated to yield
\beq\label{eqf}
{\mathring{A}}_y &=& {\mathring{\mathsf{A}}}_y \left[ 1- k^2\int_0^u \frac{du'}{1+u'+u'^2} - \int_0^u\left(\int_{u'}^1 2|\upvarphi_0|^2du''+\cdots\right) \frac{du'}{\sqrt{f(u')n(u')}}  \right]~.
\eeq
A regular solution exists since the $u'$-integral involves the $1/\sqrt{f(u')n(u')}$ term, whose eventual divergence at the horizon can be avoided, by the vanishing of the $u''$-integral at the horizon. 
The first term in Eq. (\ref{eqf}) represents the magnetic induction $\mathring{\mathsf{B}}=iq\mathring{\mathsf{A}}_y$. Therefore, a solution for the magnetic induction exists even for non-vanishing values of the uniform condensate. It implies that the lower critical magnetic field for the holographic superconductor under the Dirichlet boundary condition equals zero. The terms of Eq. (\ref{eqf})  encode the current, 
\begin{align}\label{corrent}
\langle{\mathring{\mathsf{J}}}_y\rangle &= \lim_{u\to0}\partial_u \mathring{A}_y = \mathring{\mathsf{A}}_y \left(-k^2 - 2\int_0^1  |\upvarphi_0|^2\,du+\cdots \right)~.
%
\end{align}
The second term in Eq. (\ref{corrent}) represents the supercurrent, namely, the superconducting current in a superconductor. However, in the absence of the Amp\`{e}re law, $\nabla\times {\mathsf{B}}=e^2 {\mathsf{J}}$, at the boundary, no Meissner effect exists. 
The first term  in Eq. (\ref{corrent}) can be interpreted as the bound current producing a diamagnetic current, even in the standard electromagnetic theory with no scalar field $\upvarphi_0$. 
By changing the boundary condition and imposing the (holographic) semiclassical equation $
\partial_j {\mathsf{F}}^{ij} = e^2 \langle{\mathsf{J}}^i\rangle$ 
yields \cite{Natsuume:2022kic} 
\begin{align}
\left[k^2+e^2\left(k^2+2\int_0^1 |\upvarphi_0|^2\,du\right)\right] \mathring{\mathsf{A}}_y &=0,\end{align}
implying that the magnetic potential, and consequently, the magnetic induction, equals zero. This result emulates the Ginzburg--Landau theory. To get a non-trivial solution, an external source has to be added to the holographic semiclassical equation, 
\begin{align}
\partial_j {\mathsf{F}}^{ij} = e^2 \langle{\mathsf{J}}^i\rangle + e^2 {\mathsf{J}}^i_\text{\textsc{ext}}~.\label{current4}
%
\end{align}
Eq. \eqref{current4} can be rewritten as \cite{Natsuume:2022kic}
\begin{align}
k^2 \mathring{\mathsf{A}}^y &= \upmu_m{\mathring{\mathsf{J}}}^y_\text{\textsc{ext}}~, 
\label{eq:ampere}
\end{align}
considering an effective magnetic permeability $\upmu_m =  \frac{e^2}{1+e^2}$.
Eq. \eqref{eq:ampere} is  the Amp\'{e}re law $\nabla\times {\mathsf{B}}=\upmu_m {\mathsf{J}}$. 
Hence, bound current can shift the vacuum value $\upmu_0=e^2$ to $\upmu_m$, which is related to the magnetic susceptibility $\upchi_m$ as 
\begin{subequations}
\begin{align}
\upchi_m &=\frac{\upmu_m}{\upmu_0}-1 
= -\frac{e^2}{1+e^2} <0,
%
\end{align}
\end{subequations}
constituting a diamagnetic current.
Denoting $I=\int_0^1 du\,|\upvarphi_0|^2$, the supercurrent can be addressed and the semiclassical equation becomes
\begin{align}
k^2 \mathring{\mathsf{A}}_y &= \upmu_m \left(-2I\mathring{\mathsf{A}}_y+{\mathring{\mathsf{J}}}_y^\text{ext}\right).
\end{align}
The case $I \neq 0$ implies that $
\mathring{\mathsf{A}}_y \propto e^{-x/\lambda}$, 
where $\lambda^{-1/2}=2\upmu_m I$, for $\lambda$ denoting the magnetic penetration length,  
preventing the Meissner effect in the so-called type-II superconductors. For type-II superconductors the magnetic field percolates the superconductor, and vortexes can be evinced.  Type-I superconducting materials are ideal diamagnets, and there is a critical magnetic induction above which superconductivity evanesces; the Meissner effect prevents the magnetic field from entering the superconductor. On the other hand, type-II superconductors have two critical magnetic fields, $H_{c1}$ and $H_{c2}$. The first one labels the critical magnetic field for which the first vortex can permeate the superconductor. The second one regulates situations where the vortexes are maximally stored inside the superconductor, in the sense that introducing one more magnetic flux breaks the superconductor \cite{Murugan:2016zal}. 
Nowhere near the vortex, the condensate can be constant, when the magnetic induction is small. One can then consider planar polar coordinates $ds^2=dr^2+r^2d\phi^2$ and the equation governing $A_\phi=A_\phi(u,r)$ reads  \cite{Natsuume:2022kic}
\begin{align}
\partial_u\left(\sqrt{f(u)n(u)}\partial_u A_\phi\right)+r\partial_r\left(\frac{1}{r}\partial_r A_\phi \right)-2|\upvarphi_0|^2 A_\phi=0.
%
\end{align}
The ansatz $A_\phi = U(u)R(r)$ yields 
\begin{align}
\frac{1}{U} \frac{d}{du}\left(\sqrt{f(u)n(u)}\frac{d}{du} U(u)\right)-2|\upvarphi_0|^2U(u) &= -\frac{r}{R}\frac{d}{dr} \left(\frac{1}{r}\frac{d}{dr} R(r) \right) = -\frac{1}{\uplambda^2}~,\label{2nd}
%
\end{align}
where $\uplambda$ stands for a separation constant. The second equation in (\ref{2nd}) describes the superconducting vortex, whose arbitrary solution can be written as a linear combination of the first and second kind Bessel functions, as
\begin{align}
R(r)=c_1r J_1\left(\frac{r}{\uplambda }\right)+c_2 rY_1\left(\frac{r}{\uplambda }\right),
\end{align}
with $c_1$ and $c_2$ integration constants. Asymptotically, one can realize that 
\beq
\lim_{r\to\infty}R(r)\propto\sqrt{r} e^{-r/\uplambda}.
\eeq
Thus, $\uplambda$ in the vortex case represents the magnetic penetration length. The first equation in (\ref{2nd}) can be expressed as 
\begin{subequations}
\begin{align}
\sqrt{f(u)n(u)}\partial_uU &= \int_u^1 \left(\uplambda^{-2}-2|\upvarphi_0|^2\right)U\,du', 
\end{align}
which can be integrated, leading to 
 \begin{align}
U &= \mathsf{U}\left\{ 1 + \int_0^u \frac{du'}{\sqrt{f(u')n(u')}} \int_{u'}^1 (1/\uplambda^2-2|\upvarphi_0|^2)\,du''+\cdots \right\}~.
%
\end{align}
\end{subequations}
For $u \ll 1$, the current is given by 
\beq
\langle {\mathsf{J}}_\phi \rangle &=& \frac{1}{r^2}\partial_uA_\phi.
\eeq
This result for the \clt{\gbb}\,\! equals the one for the AdS${}_4$-RN black brane \cite{Natsuume:2022kic}. 
Imposing the semiclassical equation, $\nabla^j{\mathsf{F}}_{\phi j}=e^2\langle{\mathsf{J}}_\phi\rangle$, it yields 
\begin{align}\label{cor}
\uplambda^2= \frac{1+e^2}{2e^2I}, 
\end{align}
which is analogous to the standard AdS${}_4$-RN black brane case. 

\subsection{Higher values of the magnetic field near $H_{c2}$}
Strengthening the magnetic field, more vortexes can be created, constituting a vortex  lattice engendering a supercurrent \cite{Maeda:2009vf}. The superconducting state is thoroughly  disrupted at the upper critical magnetic field $H_{c2}$ \cite{Natsuume:2022kic}. 
Different from the current induced by a magnetic field, which can be computed by the linear
response to the magnetic potential, the supercurrent consists of the dissipationless flow of the paired electrons. 
However, as in the standard holographic superconductor, there is no Maxwell equation on the AdS boundary. Therefore, the magnetic field can enter the superconductor not only at vortex cores. The holographic semiclassical equation was shown in Ref. \cite{Natsuume:2022kic} to imply the Meissner effect for the \adr. The case for the \clt{\gbb}\,\! will be  addressed. Near the upper critical magnetic field, the scalar field,  representing the condensate, is residual. Hence,  one can expand any matter field as a $\varepsilon$-series, where $\varepsilon$ denotes  the deviation parameter with respect to the critical point. The condensate and the temporal and spatial components of the magnetic potential are,  respectively, expanded as
\begin{subequations}
\begin{align}
\upvarphi(\vec{x},u) &= \varepsilon\upvarphi^{\scalebox{.65}{(1)}}+\cdots~, \\
A_t(\vec{x},u) &= A_t^{\scalebox{.65}{(0)}}+\varepsilon^2 A_t^{\scalebox{.65}{(2)}}+\cdots~, \\
A_i(\vec{x},u) &= A_i^{\scalebox{.65}{(0)}}+\varepsilon^2 A_i^{\scalebox{.65}{(2)}}+\cdots~.
%
\end{align}
\end{subequations}
At zeroth order, the static bulk equations \eqref{eq:eom_bulk} read, for the \clt{\gbb}, \begin{subequations}
\beq
\left(\sqrt{f(u)n(u)}\partial_u^2 +\partial_x^2+\partial_y^2\right)A_t^{\scalebox{.65}{(0)}}&=&0, \\
 \left[\partial_u(\sqrt{f(u)n(u)}\partial_u)+\partial_x^2+\partial_y^2\right] A_i^{\scalebox{.65}{(0)}} -\partial_i\left(\delta^{jk}\partial_jA_k^{\scalebox{.65}{(0)}}\right)&=&0, \\
 \partial_u \left(\delta^{jk}\partial_jA_k^{\scalebox{.65}{(0)}}\right)&=&0,
%
\eeq
\end{subequations}
Therefore, the Maxwell equations yield 
\begin{align}
A_t^{\scalebox{.65}{(0)}} = \mu(1-u),\qquad\quad
A_x^{\scalebox{.65}{(0)}} = 0~,\qquad\quad
A_y^{\scalebox{.65}{(0)}} = Hx~.
%
\end{align}
where $H$ denotes the critical homogeneous magnetic field and $\mu$ is, as previously, the chemical potential.
At first order, the bulk scalar equation \eqref{eq:eom_bulk} can be written as \begin{align}
\left[ -\partial_u(\sqrt{f(u)n(u)}\partial_u) + \frac{m^2}{u^4} - \frac{\mu^2(1-u)^2}{\sqrt{f(u)n(u)}} 
- \partial_x^2 - (\partial_y - iHx)^2 \right]  \upvarphi^{\scalebox{.65}{(1)}}=0~.
%
\end{align}
Using the ansatz $\upvarphi^{\scalebox{.65}{(1)}} = e^{iky}\upchi_k(x) \rho(u)$ 
yields 
\begin{subequations}
\begin{align}
\left[ -\partial_u(\sqrt{f(u)n(u)}\partial_u)+\frac{m^2}{u^4}-\frac{\mu^2(1-u)^2}{\sqrt{f(u)n(u)}} \right]\rho(u) &= -E \rho(u)~, 
\label{eq:eom_rho} \\
\left[ -\partial_x^2+H^2 \left(x-\frac{k}{H}\right)^2 \right]\upchi_k(x)  &= E \upchi_k(x)~,
\label{eq:chi}
\end{align}
\end{subequations}
where $E$ is a separation constant. By denoting 
\beq
\zeta= \sqrt{H}\left(x-\frac{k}{H}\right),
\eeq the regular solution of Eq. (\ref{eq:chi}) can be written in terms of the Hermite function, $H_n$, as
\begin{align}
\upchi_k(z) = e^{-\zeta^2/2}H_n(z)~,\end{align}
corresponding to the quantized eigenvalue $E=E_n=(2n+1)H$. 
The droplet solution yields a vortex lattice and corresponds to putting $n=0$, being consistent with 
\begin{align}
\upchi_k(x) = \exp\left[-\frac{H}{2}\left(x-\frac{k}{H}\right)^2 \right]~.
\end{align}
The general solution wits 
\begin{align}
\upvarphi^{\scalebox{.65}{(1)}} &= \rho_0(u)\Sigma(x,y)~,
\label{eq:varphi_0}
\end{align}
where 
$
\Sigma(x,y) = \int_{\mathbb{R}}  C(k) e^{iky} \upchi_k(x)\,dk,$  
and $\rho_0$ is the solution of Eq. \eqref{eq:eom_rho}, with $E=H$. 
One can obtain the vortex lattice solution by suitably choosing $C(k)$. The first order solution Eq. \eqref{eq:varphi_0} satisfies
\begin{align}
2\Im \left[\left(\upvarphi^{\scalebox{.65}{(1)}}\right)^\dagger D_i^{\scalebox{.65}{(0)}}\upvarphi^{\scalebox{.65}{(1)}} \right] = -\varepsilon_i^{~j} \partial_j|\upvarphi^{\scalebox{.65}{(1)}}|^2~,
\label{eq:bulk_current}
\end{align}
for $\varepsilon_{ij}$ denoting the 2-index Levi--Civita tensor   \cite{Maeda:2009vf}. 

Now, the Maxwell equation, in second order,  is given by
\begin{subequations}
\beq
-\left[\partial_u(\sqrt{f(u)n(u)}\partial_u) + \partial_x^2+\partial_y^2\right] A_i^{\scalebox{.65}{(2)}} + \varepsilon_i^{~j} \partial_j|\upvarphi^{\scalebox{.65}{(1)}}|^2 + \partial_i\left(\delta^{jk}\partial_j A_k^{\scalebox{.65}{(2)}}\right)&=&0~, \\
\partial_u\left(\delta^{jk}\partial_j A_k^{\scalebox{.65}{(2)}}\right)&=&0.
\label{eq:gauge}
\eeq
\end{subequations}
Eq. \eqref{eq:gauge} states that the term $\delta^{jk}\partial_j A_k^{\scalebox{.65}{(2)}}$ does not depend on $u$ and it can be chosen to vanish, as a gauge transformation  maintaining the $A_u=0$ gauge.  In momentum space, 
\begin{align}
\left(-\partial_u(\sqrt{f(u)n(u)}\partial_u)+k^2\right) \mathring{A}_i^{\scalebox{.65}{(2)}} + i\varepsilon_i^{~j} k_j {|\mathring\upvarphi^{\scalebox{.65}{(1)}}|^2}~=0,
\end{align}
where $ {|\mathring\upvarphi^{\scalebox{.65}{(1)}}|^2}$ denotes the Fourier transformation of $|\upvarphi^{\scalebox{.65}{(1)}}|^2$. 

To show the Meissner effect, it is enough to take the long-wavelength $k\to0$ limit.
One  uses the tortoise coordinate $u_\star$, defined for the \clt{\gbb}\,\!    as
\begin{align}
du_\star &:=\frac{du}{\sqrt{f(u)n(u)}},
\end{align}
or, explicitly, 
\beq
\!\!\!\!\!\!\!\!\!\!\!\!\!\!\!\!\!\!\!\!\!\!\!\!\!\!\!\!\!\!\!\!\!\!\!\!u_\star\!&\!=\!&\!\frac1{(\beta
   +2)^2 (u-1)}{\sqrt{\frac{u-1}{3 (\beta +2) u-9}}}\nonumber\\
   \!\!\!\!\!\!\!\!\!\!\!\!&\times\!&\! \left[\!(\beta \!+\!2) (u\!-\!1) ((\beta \!+\!2) u\!-\!3)\!-\!(\beta
   -1)^{3/2} \sqrt{\beta \!+\!2} \sqrt{u\!-\!1} \sqrt{\frac{3 (u\!-\!1)}{\beta\! -\!1}\!+\!u} \sinh
   ^{-1}\!\left(\frac{\sqrt{u\!-\!1}}{\sqrt{\frac{\beta \!-\!1}{\beta \!+\!2}}}\right)\right].
\eeq
 For $\beta=1$, the $u_\star\to \infty$ limit does correspond to the horizon. Then,
\begin{align}
 \left(-\frac{d^2}{du_\star^2}+ k^2 \sqrt{f(u_\star)n(u_\star)}\right)\mathring{A}_i^{\scalebox{.65}{(2)}} +i\varepsilon_i^{~j} k_j \sqrt{f(u)n(u)} {|\mathring\upvarphi^{\scalebox{.65}{(1)}}|^2} = 0. \label{gre}
%
\end{align}
Using the bulk Green's function, 
\beq
\left(-\frac{d^2}{du_\star^2} + k^2 \sqrt{f(u_\star)n(u_\star)}\right) G(u_\star,u_\star') &= \delta(u_\star-u_\star'),
\eeq
the solution of Eq. (\ref{gre}) can be formally written as
\begin{align}
\mathring{A}_i^{\scalebox{.65}{(2)}} &= \mathsf{a}_i - i\varepsilon_i^{~j} k_j\int_0^\infty  G(u_\star,u_\star') \sqrt{f(u_\star')n(u_\star')} {|\mathring\upvarphi^{\scalebox{.65}{(1)}}|^2}(u_\star')\,du_\star'.\label{aii}
\end{align}
Imposing the boundary conditions $\lim_{u_\star\to0}G(u_\star,u_\star')=0$ and $\lim_{u_\star\to\infty}\partial_\star G(u_\star,u_\star')=0$, the term $\mathsf{a}_i$ in Eq. (\ref{aii}) consists of the homogeneous solution of the PDE
\begin{align}
\left(-\frac{d^2}{d u_\star^2}+k^2\sqrt{f(u_\star)n(u_\star)}\right)\mathsf{a}_i = 0~,\label{sol1}
\end{align}
if boundary conditions regular at the horizon are imposed as well as $\lim_{u\to0}\mathsf{a}_i=\mathring{\mathsf{A}}_i^{\scalebox{.65}{(2)}}$. 
In the limit $u\to0$, the tortoise-like coordinate $u_\star$ reads
\beq
\lim_{u\to0}u_\star= -\frac{1}{(\beta +2)^2}\left[{\beta +\sqrt{\frac{\beta +2}{3-3 \beta }} (\beta -1)^{3/2} \sin
   ^{-1}\left({\sqrt{\frac{\beta +2}{\beta -1}}}\right)+2}\right].\eeq
One can construct the homogeneous solution by the $k$-expansion satisfying the boundary conditions, with $u\approx0$,  
\beq
\mathsf{a}_i\sim \mathring{\mathsf{A}}_i^{\scalebox{.65}{(2)}} (1-k^2 u+ \cdots).\label{aaa}
\eeq
The expansion in Eq. (\ref{aaa}) was obtained for the AdS${}_4$-RN black brane. For the \clt{\gbb}, one can try to emulate this result. The general solution of Eq. (\ref{sol1}), for $u\ll1$, can be written as
\beq
\mathsf{a}_i(k,\beta)&=&\left(c_1 D_{\xi_+(k,\beta)} +c_2 i^{1/2} D_{\xi_-(k,\beta)} \right)\sqrt{k}\sqrt[4]{\beta ^2+2 \beta
   -3}\left({\frac{1}{\sqrt[4]{6}}}   u+\frac{\sqrt[4]{\frac{8}{3}}}{\beta +3}\right)\label{sol11}
\eeq
where $c_1$ and $c_2$ are integration constants, and 
\beq
\xi_\pm(k,\beta)&=&\pm \frac{\sqrt{6}   k(5\beta+19)i \mp 3(\beta+3) 
 \sqrt{\beta ^2+2 \beta -3}}{6 (\beta +3) \sqrt{\beta ^2+2
   \beta -3}}
\eeq
are the order of the parabolic cylinder Weber functions $D_{\xi_\pm}(z)$. 
The profile of the solution \eqref{sol11} is depicted in what follows, for some values of $c_1$ and $c_2$.
Fig. \ref{fg1} illustrates the case $c_1=1$ and $c_2=0.1$. For these values, the field $\mathsf{a}_i$ is a monotonically increasing function of $\beta$, irrespectively of the value of $k$.  
\begin{figure} [H]
	\centering
	\includegraphics[scale=0.55]{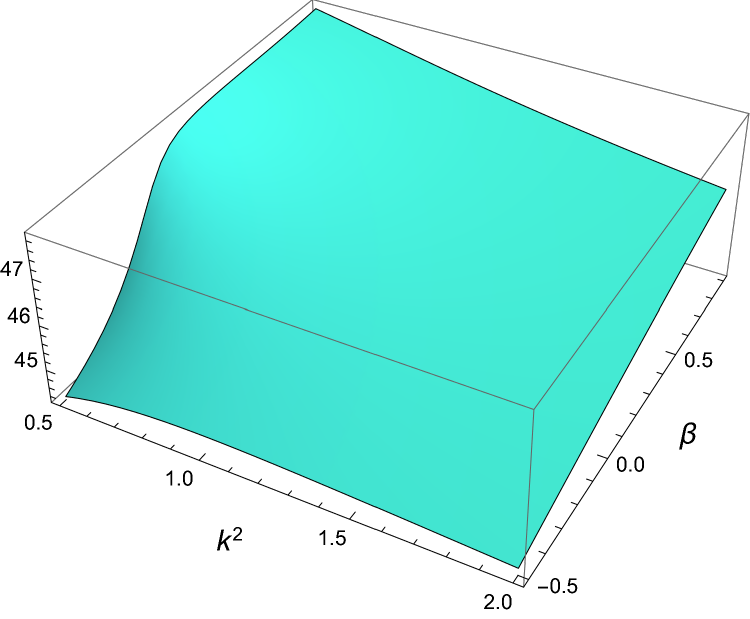}
	\caption{Plot of $\mathsf{a}_i=\mathsf{a}_i(k,\beta)$ in Eq. \eqref{sol11}, for $c_1=1$ and $c_2=0.1$. }
		\label{fg1}
\end{figure}
\noindent One can choose $c_1=10$ and $c_2=-10$, which is portrayed in 
Fig. \ref{fg2}. For these values, the field $\mathsf{a}_i$ is a monotonically increasing function of $k^2$, irrespectively of the value of $\beta$, and a monotonically increasing function of $\beta$ for $k\gtrsim 0.8$. 
\begin{figure} [H]
	\centering
	\includegraphics[scale=0.55]{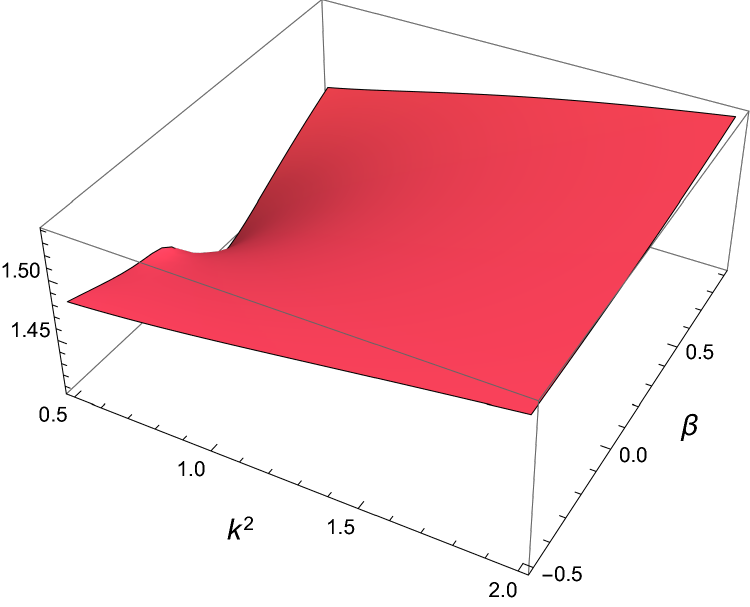}
	\caption{Plot of $\mathsf{a}_i=\mathsf{a}_i(k,\beta)$ in Eq. \eqref{sol11}, for $c_1=10$ and $c_2=-10$.}
		\label{fg2}
\end{figure}
Now, the choice $c_1=5$ and $c_2=-10$ yields 
Fig. \ref{fg3}.
\begin{figure} [H]
	\centering
	\includegraphics[scale=0.55]{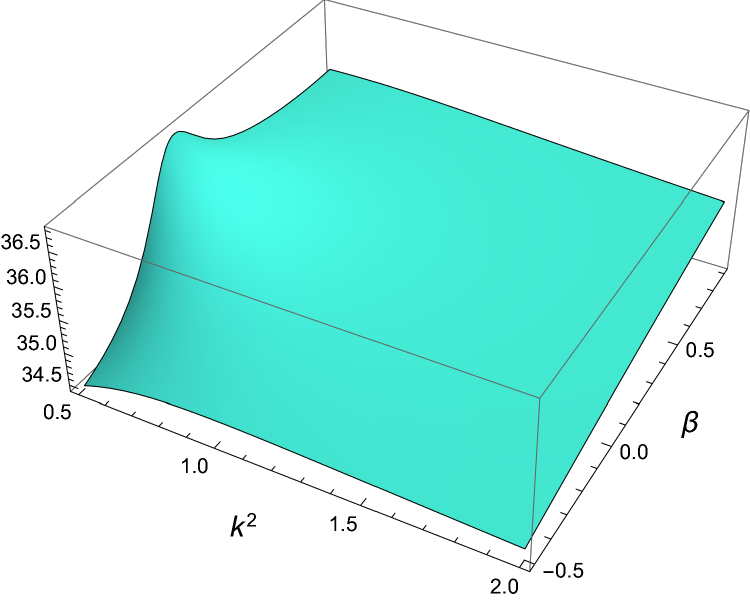}
	\caption{Plot of $\mathsf{a}_i=\mathsf{a}_i(k,\beta)$ in Eq. \eqref{sol11}, for $c_1=10$ and $c_2=-10$.}
		\label{fg3}
\end{figure}
One can construct the homogeneous solution by the $k$-expansion
\beq
\mathsf{a}_i(k,\beta) =\mathsf{a}_0(\beta) +k^2\mathsf{a}_2(\beta) +k^4\mathsf{a}_4(\beta)+\ldots,
\eeq where  the coefficients $\mathsf{a}_{2k}$, which can be read off Eq. (\ref{sol11}) and Figs. \ref{fg1} -- \ref{fg3}, depend of the parameter $\beta$ in the black brane.
When $\beta \to 1$,  one recovers the results for the \adr. 
No, realizing that since the term $i\varepsilon_i^{~j} k_j \sqrt{f(u)n(u)} {|\mathring\upvarphi^{\scalebox{.65}{(1)}}|^2}$ is proportional to $k$, it is enough to consider the Green's function at $k\approx0$, 
\begin{align}
-\partial_{u_\star}^2 G(u_\star,u_\star')  = \delta(u_\star-u_\star').
%
\end{align}
The Green's function can be obtained from two homogeneous solutions, 
\begin{align}
g_1 &=u_\star~, \qquad\qquad\qquad
g_2=1~,
\end{align}
respectively satisfying the boundary condition at the AdS boundary and the horizon. Therefore, the Green's function reads 
\begin{align}
G(u_\star,u_\star') 
= \left\{
\begin{array}{ll}
g_1(u_\star)g_2(u_\star') =u_\star' & (u_\star'<u_\star<\infty) 
\nonumber \\
g_1(u_\star')g_2(u_\star) =u_\star & (0<u_\star<u_\star') 
\nonumber
\end{array}
\right.
\end{align}
Thus, up to order $\mathcal{O}(k^3)$, one can write 
\begin{align}
\mathring{A}_i^{\scalebox{.65}{(2)}} =& \mathsf{a}_i - iu_\star\varepsilon_i^{~j} k_j \int_{u_\star}^{\infty} \sqrt{f(u_\star')n(u_\star')} {\left|\mathring\upvarphi^{\scalebox{.65}{(1)}}\right|^2} du_\star'- i\varepsilon_i^{~j} k_j \int_0^{u_\star}  u_\star' \sqrt{f(u_\star')n(u_\star')} {\left|\mathring\upvarphi^{\scalebox{.65}{(1)}}\right|^2\,du_\star'}.
%
\end{align}
The effective current is given by
\begin{align}
\langle\mathring{\mathsf{J}}_i\rangle 
&= \frac{\partial \mathsf{a}_i}{\partial{u_\star}} - \varepsilon_i^{~j} k_j \int_0^\infty du_\star'\, \sqrt{f(u_\star')n(u_\star')} {|\mathring\upvarphi^{\scalebox{.65}{(1)}}|^2}\nonumber\\
&= \left[1+\frac{1}{36} \left(1-\beta\right)\right]\mathring{\mathsf{J}}_i^n +\mathring{\mathsf{J}}_i^s~.\label{eq:current} 
%
\end{align}
where $\mathring{\mathsf{J}}_i^n=  -k^2 \mathring{\mathsf{A}}_i^{\scalebox{.65}{(2)}}$ is the bound current and the second term of Eq. \eqref{eq:current} is the supercurrent. The bound current is enhanced by the parameter $\beta(\mathfrak{q})$ controlling the AdS${}_4$  generalized black brane solutions. For $\beta\to1$, the results for the \adr\, are recovered. Therefore, we conclude that the bound current can be enhanced up to $\sim 8.2\%$ when compared to the \adr\,case, for the saturation occurring for the  lowest allowed value in the range $-1.95\lesssim \beta \leq 1$, for the extremal \clt{\gbb}. 


\section{DC conductivity of AdS${}_4$ generalized extremal branes}
\label{sec:4}
As a final result, one can use the linear response to compute the electrical DC conductivity as a function of the frequency, in the dual CFT, of the holographic superconductor in the \clt{\gbb}\,\! background.
For it, the fluctuations of the gauge potential $A_i$  in the bulk have to be evaluated. The Maxwell equations at $k\approx0$, with a stationary time dependence $e^{-i\omega t}$, can be expressed as
\beq\label{hartn}
u^4 \partial_u^2 A_i
+\left(2u^3+\frac{\left(f'(u)n(u)+n'(u)f(u)\right)}{2\sqrt{f(u)n(u)}}\right)\partial_u A_i 
+\left(\frac{\omega^2}{f^2(u)}+\frac{m^2\upphi^2(u)}{\sqrt{f(u)n(u)}}\right)A_i = 0.
\eeq
One can solve Eq. (\ref{hartn}) when ingoing wave boundary conditions at the horizon are used \cite{Hartnoll:2008vx,Son:2002sd}. 
The asymptotic behavior, for $u\to0$, takes the form
\beq
\delta A_i = \delta A_i^{\scalebox{.65}{[0]}}+  A_i^{\scalebox{.65}{[1]}}u + \mathcal{O}(u^2),
\eeq
for the dual source and expectation value for the current
being given by $\delta A_i^{\scalebox{.65}{[0]}}\sim A_i$, $ A_i^{\scalebox{.65}{[1]}}\sim \langle J^x\rangle$, whereas the electric field component reads $E_i=\lim_{u\to0}\partial_t\left(\delta A_i\right)=i\omega\delta A_i^{\scalebox{.65}{[0]}}$. In the AdS${}_4$ context of the holographic superconductor, the term $-A_i^{\scalebox{.65}{[0]}}$ can be interpreted as the superfluid velocity, whereas $A_i^{\scalebox{.65}{[1]}}$ is the supercurrent \cite{Zeng:2012xy}.
Therefore, the conductivity reads, by Ohm's law, 
\beq
\sigma(\omega) = \frac{\langle J^i\rangle}{E^i}=-i\frac{\delta A_i^{\scalebox{.65}{[1]}}}{\omega\delta A_i^{\scalebox{.65}{[0]}}}.
\eeq
In the near-boundary limit, $u\to0$, one can expand the complex charged scalar field, playing the role of the Higgs field in Ginzburg--Landau theory, as \cite{Hartnoll:2008vx}
\beq
\upphi(u) = \upphi_1 u+\upphi_2 u^2. 
\eeq
One can identify $\upphi_1$ to the source and $\langle \mathcal{O}_2\rangle\propto\upphi_2$ the condensate $\langle \mathcal{O}_2\rangle$. 
Although the range $-1.95\lesssim \beta < 1$ is formally allowed, the most relevant cases rely on realizing the \clt{\gbb}\,\! as a deformation of the \adr, and the parameter $\beta$ as a perturbation. As a consequence the values near $\beta \sim 1$ are relevant for comparing to the well-known transport coefficients associated with the \adr. 
\begin{figure} [h!]
	\centering
	\includegraphics[scale=0.55]{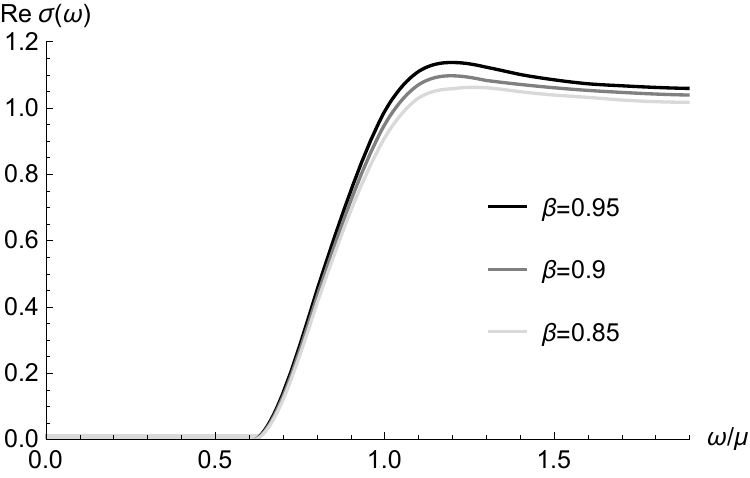}
	\caption{Real part of the DC conductivity, ${\rm Re}\,\sigma(\omega)$, at vanishing temperature for holographic superconductors with \clt{\gbb}\,\! background, as a function of the frequency-to-chemical potential ratio, for values of $\beta$ deforming the \adr. }
		\label{dois1}
\end{figure}
Fig. \ref{dois1} shows a superconducting gap in the DC conductivity for values $\omega/\mu \approx 0.62$, irrespectively the value of $\beta$. It complies with the properties of holographic superconductors, since wave states having energy smaller than the energy gap set by the order parameter, cannot be filled. From 
the quantitative point of view, the real part of the DC conductivity of the holographic superconductor in Fig. \ref{dois1} approaches the one for the \adr\, when $\beta\to1$. The lower the value of $\beta$, which controls the AdS$_4$ generalized black brane (\ref{rngen}) with metric terms (\ref{rnads41}, \ref{eq:1}), the lower the maxima 
of the real part of the DC conductivity are, and the lower its asymptotic value as a function of $\omega/\mu$. The value of $\omega/\mu$ for which the superconducting gap sets in has no significant alteration concerning the \adr, irrespectively of the value of $\beta$.  
\begin{figure} [h!]
	\centering
	\includegraphics[scale=0.55]{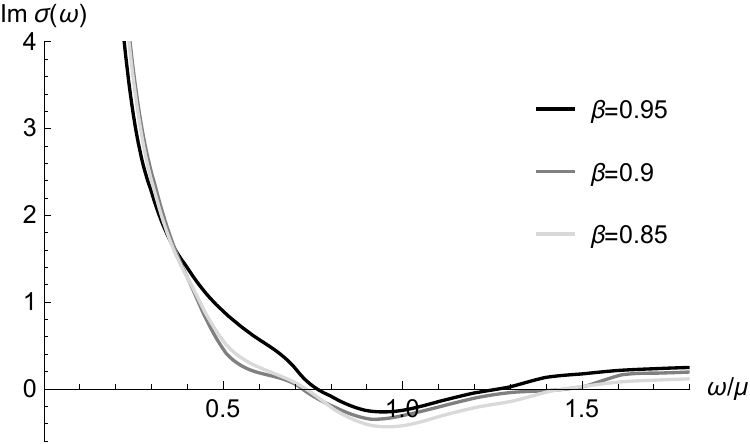}
	\caption{Imaginary part of the DC conductivity, ${\rm Im}\,\sigma(\omega)$, at vanishing temperature for holographic superconductors with \clt{\gbb}\,\! background, as a function of the frequency-to-chemical potential ratio, for values of $\beta$ deforming the \adr. }
		\label{dois2}
\end{figure}
The imaginary part of the DC conductivity of the holographic superconductor in Fig. \ref{dois2} for the \clt{\gbb}\,\! is qualitatively similar to the \adr. However, the lower the value of $\beta$,  the lower its asymptotic value as a function of $\omega/\mu$ and the higher the value of Im $\sigma(\omega)$, for fixed values of $\omega/\mu\gtrsim 0.35$.

\section{Concluding remarks and perspectives} \label{sec:5}

  \clt{\gbbs}\,\! were studied as the gravitational dual system to holographic superconductors. Transport and response coefficients, such as the shear viscosity-to-entropy density ratio, $\eta/s$, and the bulk viscosity-to-entropy density ratio, $\zeta/s$, were obtained through gravitational perturbations, using the GKPW relation and the respective Kubo's formul\ae\, for the shear and bulk viscosities. They were reported and discussed in the light of the parameter $\beta$ that governs the family of \clt{\gbbs}, \clt{whose charge $\mathfrak{q}$ is also a function of $\beta$}, as long as a second event Killing horizon exists. Maxwell equations were solved, to obtain an explicit expression for the magnetic potential involving the $\beta$-dependent chemical potential. Using the GKPW relation,  the $\eta/s$ and $\zeta/s$ ratios were derived and discussed, yielding a more strict range for $\beta$.   The holographic computation of the 
 Weyl anomaly associated with \clt{\gbbs}\,\! was implemented, with the near-boundary limit 
having an analogous result to the \adr, as expected.  The holographic superconductor in the AdS$_4$ generalized black brane background, in the probe limit, was considered and implemented with a  complex charged scalar field playing the role of the Higgs field in Ginzburg--Landau theory, minimally coupled to Einstein--Maxwell-AdS$_4$ theory. The Breitenlohner--Freedman bound
was analyzed and the equations of motion, resulting from the variational principle applied to the Higgs--Einstein--Maxwell-AdS$_4$ action, 
were solved and discussed, with applications to type-I and type-II superconductors. The case of type-II superconductors was discussed, with a vortex lattice formed by the magnetic field percolating the superconductor. The regimes of small and high values of the magnetic field were approached. We showed that the bound current strengthens by a numerical factor dependent of $\beta$, which can be up $\sim$ 8.2\% greater when compared to the AdS$_4$-RN black brane case, when the range \eqref{195}, for which $\eta/s$ is positive, is regarded. Ref. \cite{Jeong:2023las}
investigated the low-energy collective excitations in holographic superconductors with dynamical Maxwell fields at the boundary. 
Quasinormal modes (QNMs) computations corroborated the characteristic features of the Anderson--Higgs mechanism, the  Higgs modes, and the plasma frequency-gapped modes were derived, showing to be consistent with the Ginzburg--Landau theory.
Eq. (\ref{cor}) is consistent with these QNMs, particularly in the vicinity of the phase transition temperature where the probe-limit analysis is effective.

The linear response theory was also employed, to calculate the electrical DC conductivity of the holographic superconductor as a function of the frequency in the dual CFT associated with the AdS$_4$ generalized black brane background. A superconducting gap in the real part of the DC conductivity occurs for frequencies $\omega = 0.62\mu$, where $\mu$ is the chemical potential, irrespectively the value of $\beta$. For the analysis involving the DC conductivity, the range $0.85\lesssim\beta\lesssim 1$ was considered, implementing the \clt{\gbb}\,\! as a deformation of the \adr. 
For the real part of the DC conductivity, the lower the value of $\beta$, the lower its maxima are, and the lower its asymptotic value as a function of $\omega/\mu$.  We conclude that the parameter $\beta$, carrying the \clt{\gbb}\,\! charge as in Eq. (\ref{eq:7}), plotted in Fig. \ref{dois}, can fine-tune transport and response coefficients of holographic superconductors, when AdS$_4$ generalized black branes are taken into account. One can try to implement 
\clt{\gbbs}\,\! in Einstein--dilaton gravity, whose dual holographic superconductors were reported in Ref. \cite{Salvio:2012at}.

One can introduce dynamical gauge fields when Neumann boundary
conditions are imposed on the AdS$_4$ boundary. A gauge symmetry emerges with dual CFT$_3$ having a spectrum containing a massless gauge field. We can study the effects of the dynamical gauge field in the vortex lattice  configurations, emulating to \clt{\gbbs}\,\! the relevant results of Refs. \cite{Domenech:2010nf}.
where it is known to significantly affect the energetics and phase transitions
As a perspective, Ref. \cite{Kuntz:2022kcw} reported 
 a functional measure on both sides of AdS/CFT, proposing corrections to transport and response coefficients in second-order relativistic hydrodynamics. The decay rate of sound waves, the energy density, the relaxation time, the pressure, and the bulk
viscosity, as well as conformal traceless tensor fields have been shown to support relevant quantum corrections. They all achieve an imaginary component that carries the instability of
the strongly-coupled fluid flows in the CFT on the boundary. 
The latest experimental data was used to explore quantum gravity with the quark-gluon plasma. One can use the functional measure,  encoding effects of quantum gravity, to study \clt{\gbbs}\,\! and the dual holographic superconductors. 
 Analog gravity models  based upon hydrodynamics have been studied  \cite{Bilic:2021zaa}, and we expect to emulate them for \clt{\gbbs}. Besides, soft-hair excitations, in the duality between generalized  Navier–Stokes equations and soft-hairy horizons established in Ref.  \cite{Ferreira-Martins:2021cga}, may be used to probe holographic superconductors. Quantum hair can be also 
 studied in the context of \clt{\gbbs}\,\cite{Montull:2011im}.

\section*{Acknowledgements}
RdR~is grateful to The S\~ao Paulo Research Foundation FAPESP (Grant No. 2021/01089-1 and No. 2022/01734-7) and the National Council for Scientific and Technological Development -- CNPq (Grants  No. 303742/2023-2 and Grant No. 401567/2023-0), and the Coordination for the Improvement of Higher Education Personnel (CAPES-PrInt~88887.897177/2023-00), for partial financial support;  to Prof. Jorge Noronha and the Illinois Center for Advanced Studies of the Universe, University of Illinois at Urbana-Champaign, and Prof. Roberto Casadio~and DIFA, Universit\`a di Bologna,  for the hospitality.

\appendix
\section{Generalized extremal brane as an exact solution of higher-order curvature terms}\label{app1}

The parameter $\beta$ appeared as a consequence of the momentum and Hamiltonian constraint in the ADM-like protocol that led to Eq. (\ref{adm}).
However, Refs. \cite{Bueno:2016lrh,Shapiro:2015uxa} studied respectively cubic gravity and Lee-Wick gravity.  
Ref. \cite{Goroff:1985th} also addressed cubic terms involving Ricci and Riemann terms as an attempt to 2-loop quantum corrections to 4-dimensional gravity. For the 
Ricci and Einstein cubic gravity to encompass terms that are not topological, alternatively to the embedding protocol, the generalized extremal brane described by the metric (\ref{rnads4}), with metric coefficients (\ref{rnads41}, \ref{eq:1}), is an exact solution of the equations of motion coming from the action 
\beq
S = \int d^4x \sqrt{-g}\left({f_3}\left(R,R_{\mu\nu},R_{\mu\nu\rho\sigma}\right)-\frac{L^2}{4}F_{\mu\nu}F^{\mu\nu}\right)+\overbrace{\lim_{u\to 0}\int \!d^{4}x\sqrt{h}K}^{I_{\text{Gibbons--Hawking}}}+S_{\text{c.t}}\,,\label{cubic1}
\eeq
where
\beq
 {f_3} &=& \left(R-2\Uplambda_4\right)+\beta_1G_{\mu\nu}\Box R^{\mu\nu} \nonumber\\ 
&& +\beta_2\left(-\frac{7}{20}R^3+\frac{7}{5}RR_{\mu\nu}R^{\mu\nu}-\frac{7}{3}R_{\nu}^{\,\mu}R_{\rho}^{\,\nu}R_{\mu}^{\;\rho}+14R_{\mu\nu}^{\;\,\rho\sigma}R_{\rho\sigma}^{\;\,\alpha\beta}R_{\alpha\beta}^{\;\,\rho\sigma}\right.\nonumber\\
&&\left.
\qquad-4R_{\mu\nu\rho\sigma}R^{\mu\nu\rho}_{\;\,\alpha}R^{\sigma\alpha}
-\frac7{20}R_{\mu\nu\rho\sigma}R^{\mu\nu\rho\sigma}R+4R_{\mu\nu\rho\sigma}R^{\mu\rho}R^{\nu\sigma}\right)\nonumber\\
&&
+\beta_3\left(\nabla_\mu R_{\rho\sigma}\nabla^\mu R^{\rho\sigma}+\nabla_\mu R_{\rho\sigma}\nabla^\sigma R^{\mu\rho}+ \nabla_\mu R \nabla^\mu R+\nabla_\mu R_{\rho\sigma\tau\xi}\nabla^\rho R^{\mu\sigma\tau\xi}\right),	\label{cubic}
\eeq 
with Gibbons--Hawking term
\beq
\lim_{u\to 0}\int \!d^{4}x\sqrt{h}K&=&-\frac{4}{u^3 \left((1+2 \beta ) u^3+2\right)^2}{ \sqrt{-\frac{\left(3 u^7-5 u^3+2\right) \left((\beta -1) u^5-u^3+1\right)}{(1+2\beta) u^4-2}}}\nonumber
\\
&&\times u^4 \left[-32 \beta +u^2 \left(2 \beta
   +u^2 \left(56 \beta +9 (\beta -1) (2+ \beta) u^{10}+6(1+2\beta ) u^7\right.\right.\right.\nonumber\\&&\left.\left.\left.-5 \left(2 \beta ^2+\beta -5\right) u^5+24 u^3-(\beta +2) (\beta +1)
   u^2-3\right)+4\right)+8\right],
\eeq
and counterterm (c.t.) given by $\sim u^{-3}\sqrt{f(u)n(u)}.$

Using Ref. \cite{Bueno:2016xff}, the rational coefficients accompanying each one of the cubic terms make such terms neither trivial nor topological in four dimensions.  Ref. \cite{Modesto:2016ofr} showed that the 4-dimensional Lee--Wick term, whose coefficient is $\beta_1$ in the action (\ref{cubic1}, \ref{cubic}), is superrenormalizable. Ref. \cite{DeFelice:2023vmj} excluded static and spherically symmetric black holes in Einsteinian cubic gravity with unsuppressed higher-order curvature terms where the temporal
and radial metric components are equivalent to each other. Our metric 
 (\ref{rnads4}), with coefficients (\ref{rnads41}, \ref{eq:1}) evades this equivalence, deforms the AdS$_4$-Reissner--Nordstr\"om geometry and it is induced by cubic curvature terms and other higher-curvature terms beyond General Relativity. I am now studying the linear stability of the generalized extremal branes
against odd-parity perturbations. Up to now, the results I have obtained comprise the cubic terms in the action (\ref{cubic1}, \ref{cubic}), whose coefficient is $\beta_2$, having three propagating degrees of freedom in the odd-parity sector. One dynamical perturbation behaves as a soft  ghost mode. It is worth mentioning that General Relativity has one dynamical perturbation. Therefore we want to explore this soft ghost mode better, also using the results in Ref. \cite{Li:2017txk}.  
About the terms in the action (\ref{cubic1}, \ref{cubic}), whose coefficient is $\beta_3$, their renormalizability must still be explored, which is still an intricate task. Therefore, in this setup, the parameter $\beta = h(\beta_1, \beta_2, \beta_3)$, where $h$ is a rational function, is a rational combination of the coefficients $\beta_1, \beta_2$, and  $\beta_3$, respectively encoding quantum gravity/stringy effects beyond General Relativity. To extend Ref. \cite{Goroff:1985th}  as an attempt to identify cubic gravity and the terms accompanying the coefficient $\beta_3$ in the action (\ref{cubic1}, \ref{cubic}) as 2-loop quantum corrections to 4-dimensional gravity, one must further investigate the terms in the action (\ref{cubic}), whose coefficient is $\beta_3$. The final answer remains unknown up to now. 

\bibliography{bibliografia}
\bibliographystyle{iopart-num}

\end{document}